# Global Drought Escalation Since 2014: Multivariate Insights for Ecosystem Resilience and Water Security

Qianqian Han[1], Yijian Zeng[1], Bob Su[1,*]

[1]Faculty of Geo-Information Science and Earth Observation (ITC), University of Twente; Enschede, 7522 NH, the Netherlands

*Corresponding author: Bob Su (z.su@utwente.nl)

**Abstract:** How is climate change altering the global patterns, frequency, and intensity of drought? To investigate this, we analyzed global drought events from 2000–2020 using a composite of hydrological indices (Σ(ET–P), SSMI, and GRACE-DSI). Our results reveal that near-normal to severe droughts became increasingly widespread after 2014. Unexpectedly, extreme, 100-year return period droughts occurred not only in semi-arid zones but also in high-latitude and typically wet regions. The frequency of these rare events increased globally, with durations exceeding 100 days in hotspots like Eurasia. Ecosystem case studies further showed divergent vulnerabilities in carbon and water cycling. These findings underscore an urgent need for region-specific adaptation strategies to enhance resilience against increasingly extreme droughts.

**Main Text:** Drought is a common natural event (*1-4*), and is a significant threat that has a considerable impact on ecosystems (*5*). Impacts of drought include devastated crops, famine and conflicts and wars (*6*). Unlike many natural disasters, such as floods or hurricanes, droughts are similar to a disease in the sense that they start before any noticeable symptoms appear (*7*). Climate projections indicate that drought occurrences will intensify in frequency, severity, and duration in future decades, driven by climate change (*8*). The increasing frequency and intensity of droughts underscore the urgent need to understand their implications on ecosystems (*9*), because droughts can affect the health, structure, and diversity of ecosystems (*9, 10*).

Meteorological indices, while widely used for drought monitoring, often lack sensitivity to regional ecological responses and hydrological feedbacks (*11, 12*). Ground-based observation networks are spatially sparse and temporally inconsistent, particularly in data-scarce regions (*13*). Moreover, many traditional metrics focus primarily on climate inputs such as precipitation, without considering the complex feedbacks between soil moisture, vegetation activity, and atmospheric evaporative demand (*14*). Additionally, they do not fully account for subsurface hydrological processes or ecosystem functioning, making it difficult to capture the full extent and ecological consequences of drought. These limitations hinder a holistic understanding of drought dynamics, particularly the complex interplay between water availability, ecosystem function, and climate feedbacks.

To overcome these limitations, remote sensing and reanalysis data offer spatially continuous and ecologically relevant insights. Satellite observations enable high-resolution monitoring of vegetation stress (e.g., via NDVI, EVI, SIF anomalies) and land surface temperature, providing early drought warning signals (*15*). Soil moisture retrievals (e.g., from SMOS, SMAP, AMSR-E) allow precise assessment of water availability critical in arid regions (*16*). GRACE satellite gravimetry uniquely captures total terrestrial water storage anomalies, including groundwater, revealing subsurface drought impacts invisible to surface measurements (*17, 18*). Operational systems like the Global Drought Observatory and U.S. Drought Monitor integrate such data streams to improve drought assessment (*19, 20*).

Building on these capabilities, multi-variable drought indicators have emerged to better characterize the distinct types and cascading effects of drought (meteorological, agricultural, hydrological, ecological), as demonstrated by work highlighting sequential propagation through these systems and its value for early warning (*21*). The importance of integrated drought indices that consider both water availability and ecosystem function was emphasized (*21, 22*). For example, the Vegetation Drought Response Index and the Evaporative Stress Index have been used to monitor vegetation stress and land-atmosphere coupling during drought events (*23, 24*).

Despite these advances, a critical challenge remains: fully integrating direct measures of ecosystem function and carbon cycling with hydrological variables to understand the ecohydrological coupling (the dynamic interactions between hydrological processes and ecosystem function) during drought. Quantifying how vegetation physiology, carbon uptake (GPP), and land-atmosphere energy/water exchanges respond to water stress is essential, yet difficult. There is therefore an increasing urgency to evaluate drought through this integrated ecohydrological lens to predict impacts on ecosystem health, carbon sequestration, and climate feedbacks.

To address this need for integrated ecohydrological assessment, this study leverages multiple satellite and model-based variables capturing water deficits, storage, availability, and ecosystem function. Specifically, we conduct a comprehensive global analysis (2000-2020) integrating: 1) cumulative water deficit dynamics ($\Sigma(ET–P)$), 2) soil moisture (SM) as a key indicator of plant-available water, 3) terrestrial water storage anomalies (GRACE) representing total subsurface reserves, 4) gross primary production (GPP) z-scores to quantify ecosystem productivity responses, and 5) energy and water flux partitioning ratios (LE/Rn, P/Rn) to assess climate regulation and efficiency. This multi-faceted approach enables us to: 1) characterize drought extent, severity, duration, and hotspots; 2) elucidate regional ecosystem vulnerabilities through GPP responses; and 3) evaluate the regulatory effects of energy and water availability on drought impacts via flux partitioning analysis across climate zones. Ultimately, this work provides critical insights to support more effective, regionally tailored drought mitigation and ecological management strategies under changing climatic conditions.

# Results

**Drought severity**

**Spatial distribution and temporal change of drought return periods**

To assess global drought severity, we analyzed three hydrological indicators: cumulative precipitation deficit ($\Sigma(ET-P)$), Standardized Soil Moisture Index (SSMI), and GRACE-DSI (total water storage). Return periods were classified into five categories (5–10, 10–25, 25–50, 50–100, >100 years) using a consistent methodology (Materials and Methods).

Spatially, long-term mean return periods (Figs. 1a, S2a, S3a) reveal that near-normal/abnormal droughts (5–25-year return periods) dominate global land areas. In contrast, severe-to-extreme droughts (>25-year return periods) are spatially limited, appearing only as localized hotspots. $\Sigma(ET-P)$ highlights hotspots across high-latitude zones, humid US regions, and semi-arid areas (Fig. 1a). SSMI shows persistent dryness (return periods <5 years) in hyper-arid regions (e.g., Sahara, Australia), with severe droughts confined to isolated pockets (Fig. S2a). GRACE-DSI similarly indicates widespread moderate conditions but reveals scattered extremes (e.g., northern Africa, southern South America; Fig. S3a), suggesting acute water storage deficits.

Temporally, daily/monthly area fractions (Figs. 1b, S2b, S3b) confirm that 5–10-year droughts dominate globally. $\Sigma(ET-P)$ and SSMI exhibit strong seasonality, peaking in June–August due to Northern Hemisphere summer dryness. Severe droughts (>25-year) are episodic: $\Sigma(ET-P)$ shows winter peaks in >100-year events (>4% coverage), while SSMI peaks occur in 2010, 2016, and 2020. GRACE-DSI fractions are smaller (2–5% for 5–10-year class) but show intensified groundwater droughts in 2019–2020 (rising 10–50-year categories). Critically, all indicators show an upward trend in drought-affected area across severity levels after 2015, indicating accelerated global drought intensification. Corresponding continent-level results are provided in Figs. S4, S5, and S6 respectively.

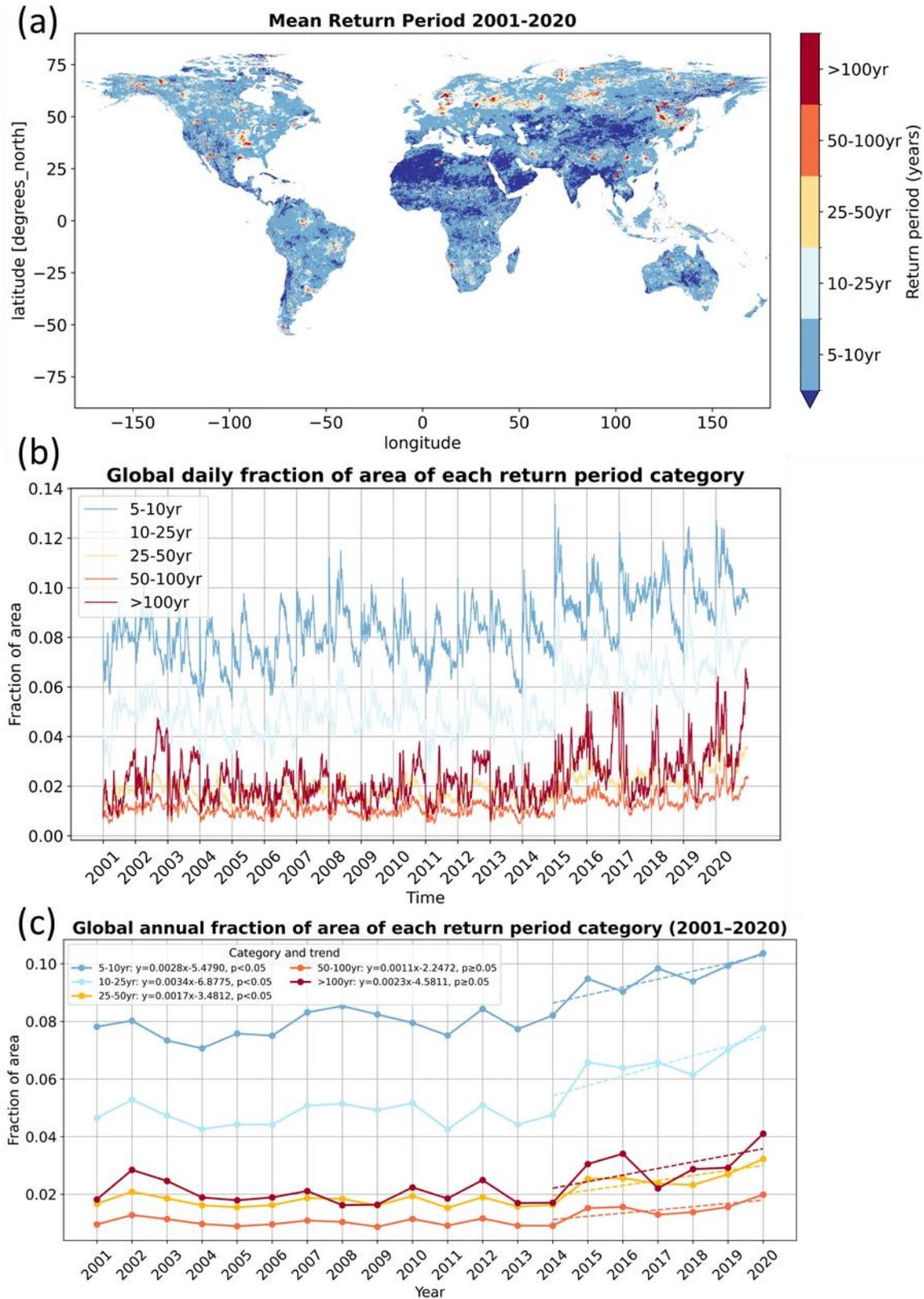

**Fig. 1.** (a) Global spatial distribution of mean drought return period based on Σ(ET-P) from 2001 to 2020, (b) Global daily fraction of area for each return period category during 2001-2020, (c)

Global annual fraction of area for each return period category during 2001-2020 based on Σ(ET–P).

## Frequency and duration based on annual cumulative ET-P

For each return period category, we calculated probability density functions (PDFs) of annual drought-day counts across continents (2001-2020 average; Fig. S15). PDFs were computed for six continents using logarithmic scaling to enhance comparison.

Key frequency patterns emerge: 1) [5–10)-year and [10–25)-year categories: Symmetric distributions peak at 20–40 drought days/year (Figs. S15a,b). 2) [25–50)-year category: Distinct leftward shift reflects reduced drought persistence, with continental divergences intensifying (Fig. S15c). Europe and South America show narrow peaks (stable drought regimes), while Oceania exhibits right-skewed distributions (high variability). 3) [50–100)-year category: Continued left shift confirms rarer droughts yield fewer annual drought days (Fig. S15d). 4) [100,+∞)-year category: Remarkable reversal toward higher drought-day counts indicates extreme droughts break the rarity-persistence relationship through prolonged atmospheric anomalies (Fig. S15e).

Drought duration analysis reveals complementary patterns (Fig. 2): 1) Progressive shortening: From [5–10)-year to [50–100]-year categories, mean/max durations decrease globally (e.g., 80–120 days to 40–60 days maxima). 2) Extreme-category divergence: >100-year events show disproportionately long maxima (>100–140 days) concentrated in high latitudes where polar amplification enhances drought persistence. Mean durations remain low (<15 days) except in hotspots (central Eurasia, eastern Russia).

In summary, the PDF reversal ([100,+∞)-year) and duration extremes spatially co-locate, confirming that the rarest droughts impose compound risks through both frequent annual occurrence and extended persistence – particularly in climate-sensitive regions.

## Accelerating Global Drought Trends

Robust expansion in drought-affected areas (2001–2020) is evident across all severity levels (Figs. 1b, S3b, 1c): 1) Significant increases ($p<0.05$): 5–10-yr: $+0.0028$ $yr^{-1}$; 10–25-yr: $+0.0034$ $yr^{-1}$; 25–50-yr: $+0.0017$ $yr^{-1}$; 2) Post-2014 intensification: 100-yr events: +54% vs. 2001–2014 baseline; Peaked during 2015–2016 El Niño (GRACE-confirmed (*25*)).

Regional contributions differ by severity: Africa/Asia dominated the expansion of moderate drought, whereas Asia, Europe, and South America led increases in extreme drought.

This aligns with global assessments(*25, 26*) attributing drought expansion to: 1) Warming-enhanced evaporative demand (58% of post-2018 increase(*26*)); 2) Intensified El Niño episodes, 3) Background climate change.

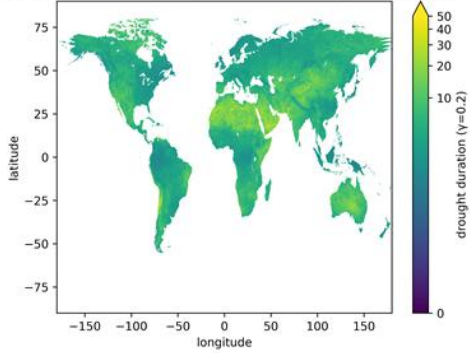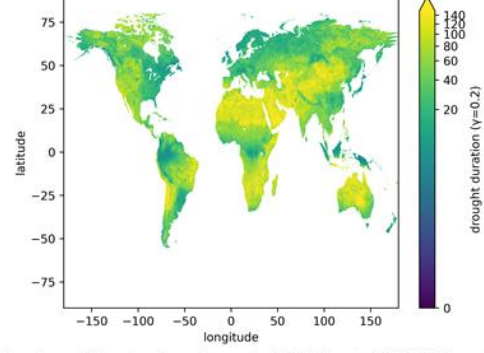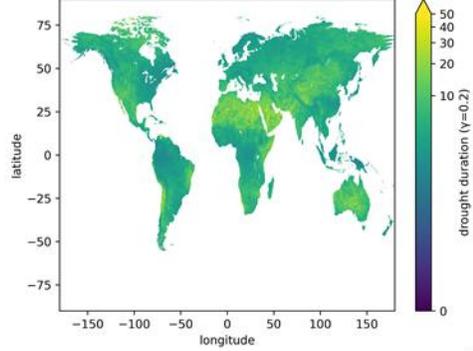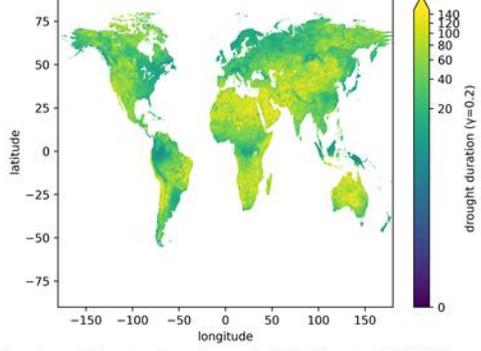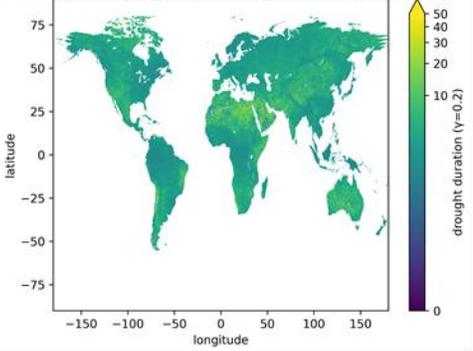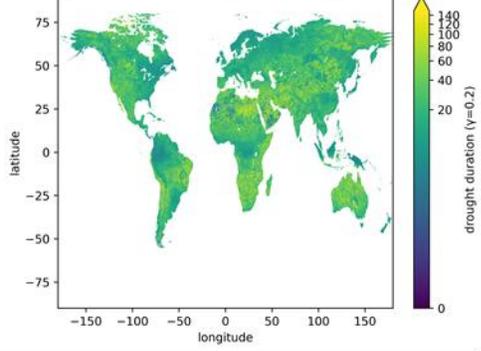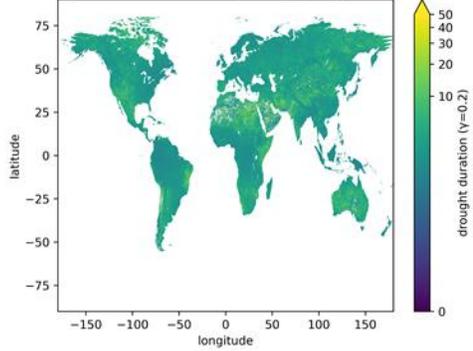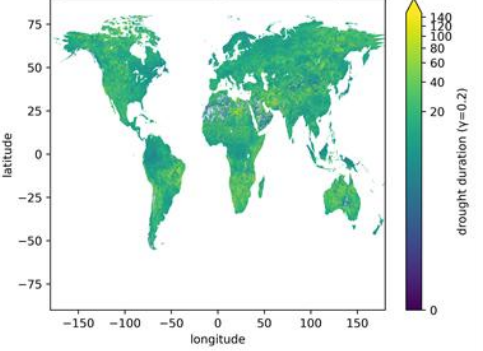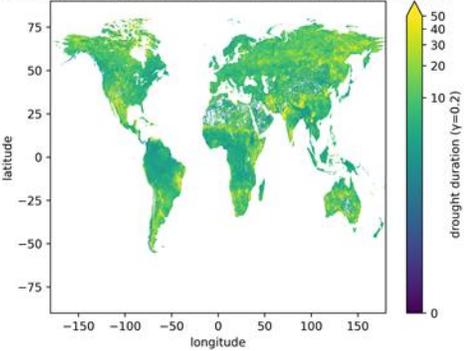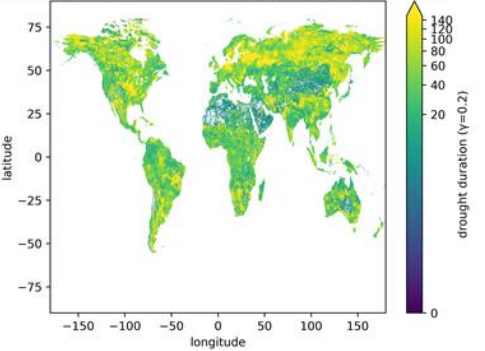

**Fig. 2.** Spatial patterns of mean (left column) and maximum (right column) drought duration (days) from 2001 to 2020, calculated for each return period category. For each grid cell, drought duration was calculated annually, then aggregated as the multi-year mean or maximum value. Color indicates the number of days a drought persisted.

## Water-Energy climatology

Given that ecosystems respond differently to droughts under water- vs. energy-limited regimes, partitioning the land surface using LE/Rn and P/Rn ratios clarifies region-specific drivers. Our global climatology (Fig. 3a) identifies four regimes: 1) Water-limited (green): High LE/Rn, low P/Rn - ET demand exceeds rainfall supply. 2) Energy-limited (magenta): High P/Rn, low LE/Rn - Radiation constrains ET despite ample rain. 3) Co-abundant (brown): High LE/Rn & P/Rn - Synergy of moisture/energy fuels strong ET. 4) Hyper-arid (gray): Low LE/Rn & P/Rn - Scarcity of both water and energy.

Spatially, co-abundant regimes dominate humid tropics (Amazon, Congo, SE Asia) and SE US (Fig. 3a). Water-limited zones flank mountain ranges (Andes, Rockies), while hyper-arid conditions prevail in subtropical deserts (Sahara, Australia).

Köppen–Geiger climate zones align with these regimes (Fig. 3b): 1) Tropical/humid temperate zones (A, Cw, Cf) are predominantly co-abundant. 2) Deserts (BW) cluster in hyper-arid zones; polar climates (E) occupy this sector but spill elsewhere during seasonal melt. 3) Cold continental zones (Dw, Df) span multiple bins due to pronounced seasonality in radiation and moisture.

Hotspot analysis reveals distinct drought drivers: 1) Water limitation drives Central Asia, NW China, and S. Africa droughts (precipitation deficits). 2) Energy limitation triggers Alaska and NE China droughts (radiative constraints suppress ET). 3) Dual drivers in Siberia and Europe reflect compound stress from rain deficits and energy imbalances.

Critically, extreme droughts may defy climatology: While the Netherlands (mean energy-limited) experienced 2018 drought from precipitation anomalies, the Amazon (co-abundant) faced stress during El Niño-induced rainfall reductions. Thus, while water-energy climatology defines background constraints, event-scale droughts arise from temporary imbalances in precipitation and/or energy fluxes.

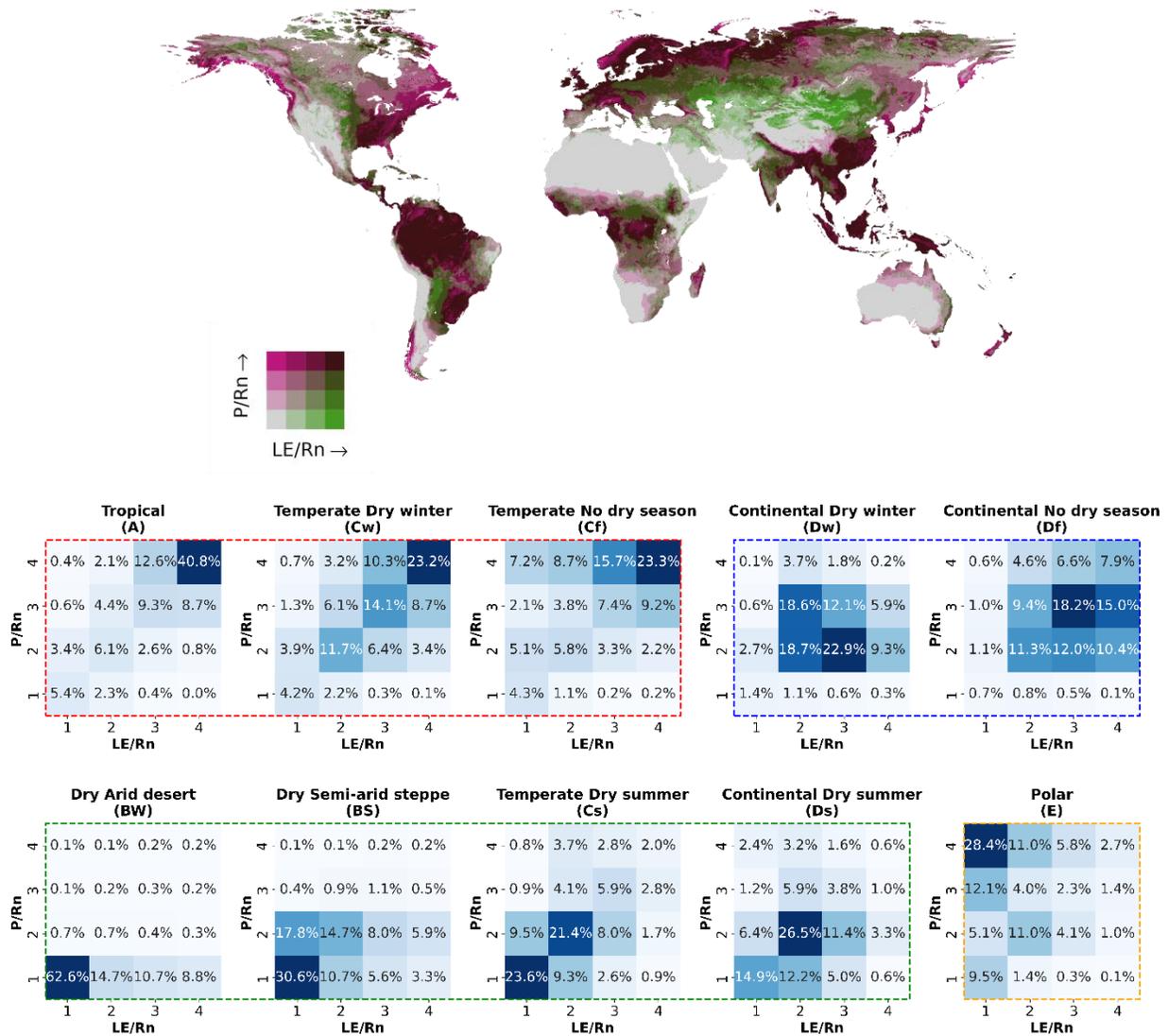

**Fig. 3.** a) Global Land Energy–Water Partitioning (LE/Rn ~ P/Rn). b) Pixel Fractions by Köppen–Geiger Climate Zones in 2000-2020

# GPP extremes

*Analysis Framework*

To identify climate drivers of terrestrial productivity extremes, we analyzed negative (lowest percentile) and positive (highest percentile) GPP anomalies during 2001–2020. For each grid cell, we: 1) Identified extreme-GPP years, 2) Extracted concurrent anomalies in temperature (T), precipitation (P), and precipitation deficit ($\Sigma(ET–P)$), and 3) Mapped composites in T~P and T~$\Sigma(ET–P)$ space (Fig. 4).

This approach reveals event-scale climate conditions during productivity crises or peaks - distinct from background climatology.

*Negative Extremes (Low Productivity)*

This category includes high-latitudes (Figs. 4 a1, a2), which cluster in cool-wet and cool-dry quadrants (blue/purple) where energy limitation dominates - low temperatures suppress photosynthesis regardless of moisture. Elsewhere, two patterns emerge: (1) Water stress—warm-dry anomalies (red in T~P) coincident with positive Σ(ET–P), where moisture deficit overwhelms ecosystems; and (2) Energy-water compound stress—warm anomalies with negative Σ(ET–P) (red in T~P; green in T~Σ(ET–P)), coincident with cloudier conditions that reduces radiation while high temperatures raise evaporative demand. Table S2 lists hotspot examples, revealing that low-GPP events are driven by region-specific climate stressors rather than mean climatology.

*Positive Extremes (High Productivity)*

This set includes the high-latitudes (Figs. 4 b1, b2), which are overwhelmingly warm-dry (red in both panels) under a temperature-control regime: warming lifts energy limitation and boosts GPP, largely independent of moisture status. In drylands (North Africa, Central Asia), responses split into: (1) water-driven: high P with negative Σ(ET–P) (unstressed vegetation), and (2) radiation-driven: high T with positive Σ(ET–P), indicating abundant moisture plus clear skies synergistically raise GPP. Seasonally dry regions (Southern Africa, Australia, southwestern North America, India) appear cool-wet in T~P (blue) but bifurcate in T~Σ(ET–P): a water-driven branch with negative Σ(ET–P) where moisture relieves limitation, and an energy-driven branch with positive Σ(ET–P) under cool, clear conditions where high radiation unlocks evaporative potential.

In summary, GPP extremes arise when climate anomalies cross functional thresholds. Negative extremes reflect (i) energy scarcity (high latitudes), (ii) moisture deficit (drylands), or (iii) compound stress from cloud-reduced radiation coupled with high evaporative demand. Positive extremes arise from (i) relief of energy limitation (high latitudes), (ii) optimal water-energy coupling (drylands), or (iii) synergistic moisture-radiation episodes in monsoonal regions. This critically implies that drought impacts manifest through regionally distinct climate pathways that override background climatology.

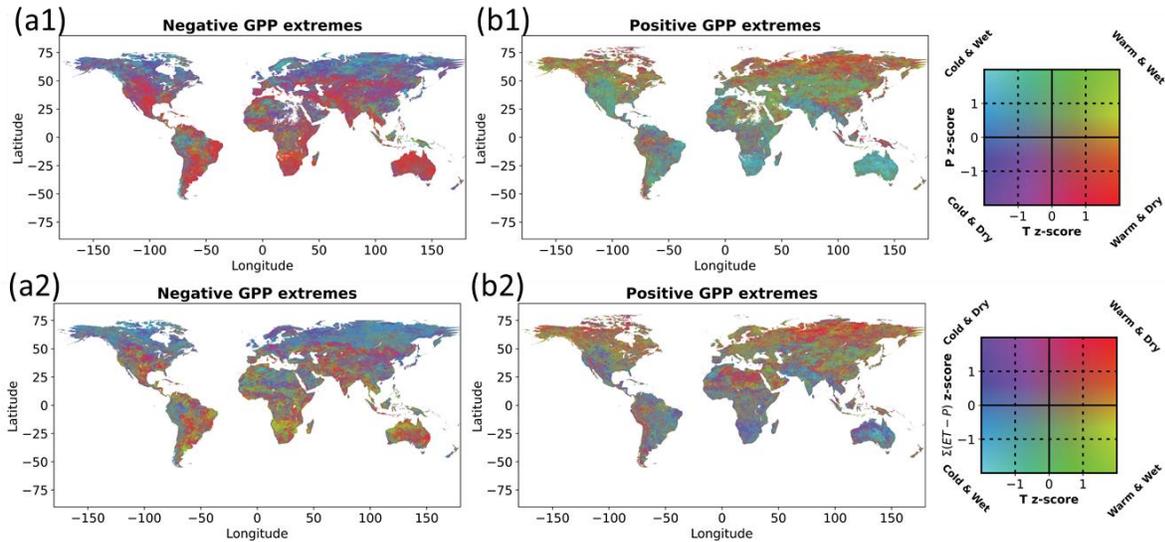

**Fig. 4.** Climatic drivers of GPP extremes: (a1) T~P z-score during lowest-decile GPP years; (a2) T~Σ(ET–P) for same; (b1) T~P during highest-decile GPP years; (b2) T~Σ(ET–P) for same. Anomalies are z-scores relative to 2001–2020 baseline.

# Discussion: Spatiotemporal patterns and drivers of extreme drought

## 1. Hotspot Attribution & Event Drivers

Our hotspot analysis (Fig. 1a) identifies regions experiencing extreme droughts (≥50-year return periods) during 2001–2020. For each hotspot, we determined the peak severity year (Fig. 5) by selecting the maximum annual median return period. Temporal frequency of extremes (Fig. S13) reveals most hotspots reflect single exceptional events, though Central Asia/southern Africa show recurrent extremes.

Key event-driver linkages emerge: Alaska (2004)—atmospheric ridging reduced snowpack, elevating wildfire risk (*27, 28*); Central U.S. (2012)—precipitation deficits plus record heat triggered agricultural collapse (*29*); Northern Europe (2018)—persistent anticyclones and an NAO phase shift produced heatwaves and rainfall shortages (*30*); Siberia—2012 anticyclonic moisture divergence reduced precipitation (*31*), whereas 2020 stratospheric polar-vortex anomalies drove record warmth (*32*); East Asia (2014) —tripolar precipitation anomalies induced regional drying (*33*); Southern Africa (2013/2017) —oceanic forcings caused record wheat-yield losses (*34-36*); Amazon (2015–2016) —El Niño disrupted convection, leading to wet-season rainfall deficits (*37, 38*).

In conclusion, extreme droughts arise from regionally specific climate anomalies - atmospheric blocking, oceanic modes, and moisture transport disruptions being recurrent drivers.

## 2. Duration of Extreme Droughts

During peak severity years (Fig. 5), extreme drought persistence ([100,+∞)-year days) exhibits strong hydroclimatic gradients (Fig. 6): highest persistence (200–300 days) occur in Southern Africa, central Australia as a result of background aridity amplifies impacts; intermediate (50–150 days) in Central Asia, U.S. Great Plains; and shortest (<50 days) in high-latitude regions (Alaska, Scandinavia) because seasonal constraints limit duration. The monthly frequency is shown in Fig. S14.

This demonstrates that drought duration is modulated by: background climate (aridity); seasonal circulation anomalies; and geographic location (e.g., latitude).

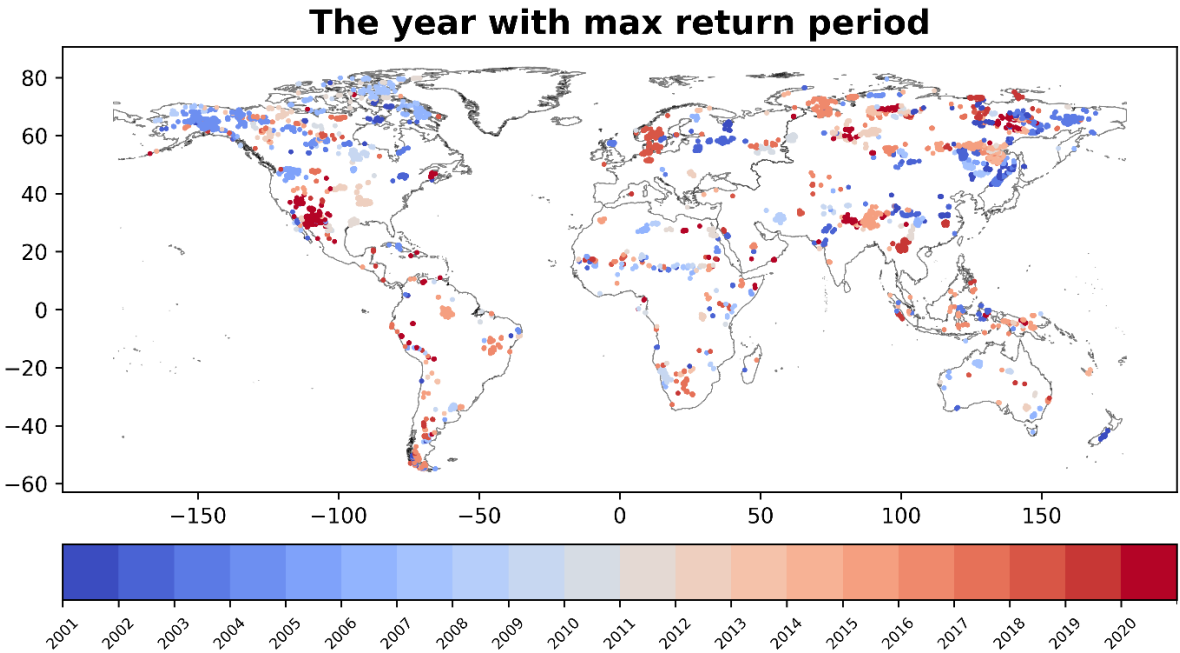

**Fig. 5.** In the hotspot regions, the year with max return period.

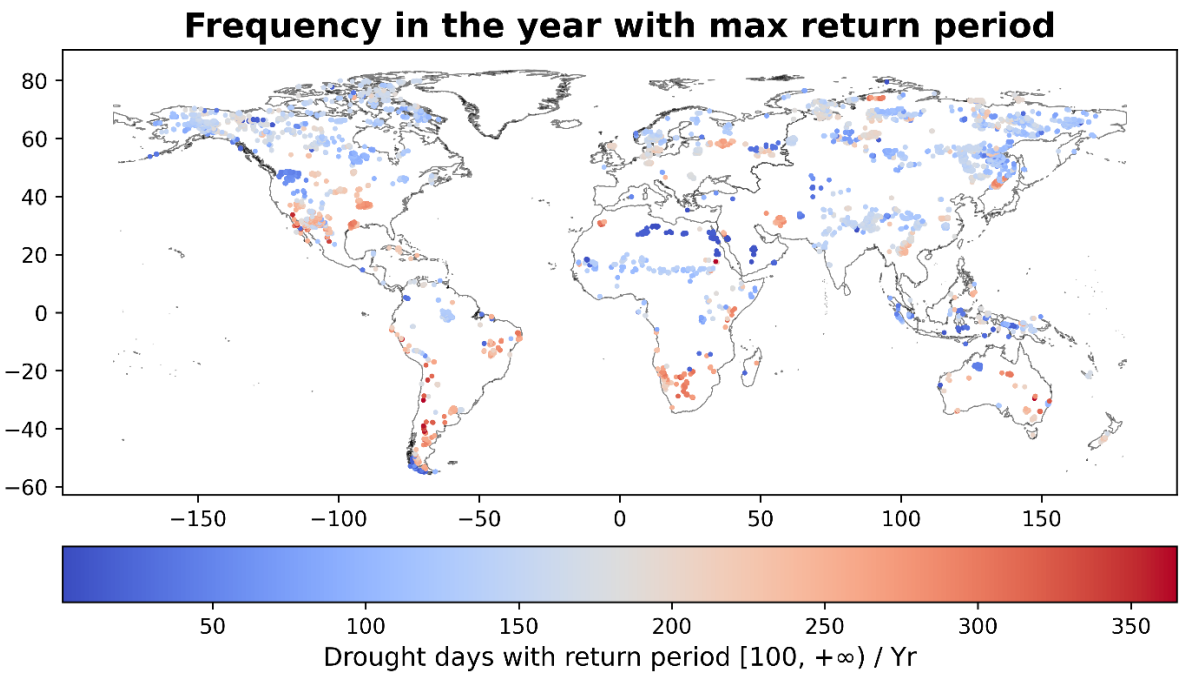

**Fig. 6.** The frequency of category return period [100,+∞) in the year with max return period

**3. Complementary Indicators Reveal Drought Propagation (Σ(ET–P), SSMI, GRACE-DSI)**

All metrics consistently identify arid/semi-arid zones as drought-prone, but divergence in extreme droughts reflects hydrological processes (Table S3). These metrics capture sequential drought stages - from rapid surface imbalances (Σ(ET–P)) to soil moisture stress (SSMI) and to deep storage deficits (GRACE-DSI). Multi-metric synergy is essential for comprehensive drought monitoring.

## 4. Ecohydrological Context of Drought Impacts

Long-term Σ(ET–P) climatology (Fig. 7) reveals: strong deficits in hyper-arid regions (>300 mm); moderate deficits in temperate zones (50–150 mm); and near-balance in tropics/boreal regions.

Notably, droughts are increasingly affecting humid regions. In the Netherlands (2018), an energy-limited climatology was overridden by a precipitation deficit, producing substantial agricultural losses; in the Amazon (2015–2016), El Niño disrupted the water-energy balance, threatening biodiversity. These cases show that while background climatology sets the operating envelope, extreme events arise when anomalies drive the system out of balance.

## 5. Addressing Key Challenges

Our framework addresses three key limitations in drought monitoring: (1) Spatiotemporal granularity—daily-resolved Σ(ET–P) reveals post-2014 intensification and discrete hotspots that annual indices miss; (2) Multidimensional assessment—combined metrics trace drought propagation across hydrological compartments; and (3) Ecohydrological linkages—GPP extremes quantify productivity responses to water- and energy- stresses (Fig. 4). Nonetheless potential soil moisture overestimation in northern China/Mongolia (due to higher organic matter vs. Sahara) may inflate LE estimates in these regions - findings here require cautious interpretation.

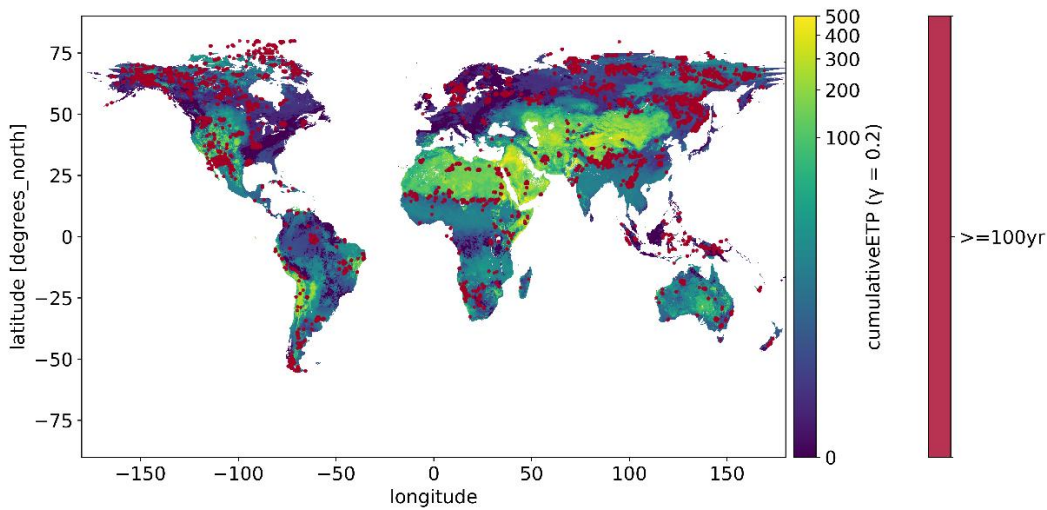

**Fig. 7.** Long-term cumulative ET-P climatology (mean of $\sum_{2000}^{2020}(ET - P)$). Color indicates cumulative ET–P in mm, with color scale nonlinearly stretched using PowerNorm (γ = 0.2).

**Funding:** The research presented in this paper was funded in part by:
The China Scholarship Council grant 202004910427 (QH)
NSO New Earth Observation Data for Water grant KNW23002 (YZ).
SURF grant EINF-6614 (YZ)
SURF grant EINF-12364 (YZ)
The Netherlands Organisation for Scientific Research (NWO) KIC, WUNDER grant KICH1.LWV02.20.004 (ZS, YZ)
Netherlands eScience Center, EcoExtreML grant 525 27020G07 (ZS, YZ).

**Author contributions:**

    Conceptualization: ZS, YZ, QH

Methodology: QH, ZS, YZ

Investigation: QH, ZS, YZ

Visualization: QH

Funding acquisition: ZS, YZ

Project administration: ZS, YZ

Supervision: ZS, YZ

Writing – original draft: QH

Writing – review & editing: YZ, ZS

**Competing interests:** Authors declare that they have no competing interests.

**Data and materials availability:** All data are available in the main text or the supplementary materials. Code is available on request.

**Acknowledgments:** We are grateful to the NASA science team members who kindly provided clarification and support via email for GRACE and IMERG data. AI-assisted technologies (ChatGPT, OpenAI) were used to improve the clarity of the language, and all content was reviewed and approved by the authors.

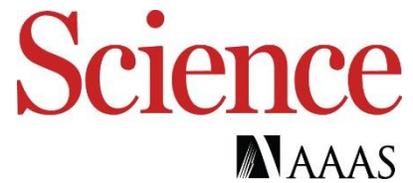

Supplementary Materials for

**Global Drought Escalation Since 2014: Multivariate Insights for Ecosystem Resilience and Water Security**

Qianqian Han, Yijian Zeng, Bob Su

Corresponding author: Bob Su (z.su@utwente.nl)

**The PDF file includes:**

    Materials and Methods
    Supplementary Text
    Figs. S1 to S15
    Tables S1 to S2
    References

**Materials and Methods**

## Datasets

Gridded hourly latent flux (can be converted to evapotranspiration), gross primary production, net radiation (Rn) from March 2000 to December 2020 were obtained from FluxHourly (*1*), with a spatial resolution of 9 km. The evapotranspiration was converted from latent heat flux. LE, ET and Rn was aggregated to daily/annual and GPP to annual values.

The precipitation, in mm/day, was obtained from IMERG, with a spatial resolution of 0.1 degree (*2*).

Air temperature (T) was from hourly ERA5-Land data, developed by the European Centre for Medium-Range Weather Forecasts (ECMWF), with a spatial resolution of 9 km. T was aggregated to daily values.

SM is from GSSM1km, with daily and 1 km resolution (*3*).

GRACE Mascon total water storage anomalies (TWSA) is from April 2002 to December 2020, with monthly and 0.5 degree resolution and version JPL RL06.3_v04 (*4*).

## Key Metrics and Calculations

### 1. Cumulative precipitation deficit Σ (ET-P)

Definition: Annual cumulative sum of daily (ET – P) from January 1–December 31 (2001–2020); Data: ET (FluxHourly), P (IMERG); Processing: Negative values set to zero; 2000 excluded due to incomplete ET data; Rationale: Σ (ET-P) reflects root-zone moisture depletion more accurately than reference evapotranspiration (ET₀).

### 2. Standardized Soil Moisture Index (SSMI)

Calculation:

$$SSMI_i = \frac{SM_i - \mu_i}{\sigma_i} \qquad (1)$$

where $SM_i$ = monthly SM (aggregated from GSSM1km daily), $\mu_i$ and $\sigma_i$ = climatological mean/standard deviation for month *i* (2000–2020). The SSMI is a dimensionless quantity that detects both drought and abnormally wet events, but we only focus on drought in this research, with drought threshold: SSMI < 0.

### 3. GRACE Drought Severity Index (GRACE-DSI)

Calculation (*5*):

$$GRACE - DSI_{i,j} = \frac{TWSA_{i,j} - \overline{TWSA_j}}{\sigma_j} \qquad (2)$$

where $TWSA_j$ and $\sigma_j$ = mean/SD of monthly climatology (2002–2020).
The GRACE-DSI is a dimensionless quantity that detects both drought and abnormally wet events, but we only focus on drought in this research, with GRACE-DSI < 0.

## 4. Water-Energy Partitioning (LE/Rn & P/Rn)

Calculation: Annual LE/Rn and P/Rn ratios averaged over 2000–2020
Data: LE (FluxHourly), Rn (FluxHourly), P (IMERG)
Purpose: Quantify energy vs. water limitation (analogous to Budyko framework) (*6*).

## 5. GPP Extremes Analysis

GPP was used to detect extremes and we presented its climatic drivers during extremes (*7*). Firstly hourly GPP was aggregated to yearly (gC/m$^2$/yr) to calculate GPP z-score for 2001-2020.

$$GPP - zscore_i = \frac{GPP_i - \mu}{\sigma} \qquad (3)$$

Where $\mu$, $\sigma$ = 2001–2020 mean/SD of annual GPP.
Extreme thresholds: Negative: 1st–10th percentile; Positive: 90th–99th percentile;
Climate driver integration: Monthly T/P anomalies detrended and standardized; Weighted climate indices derived via regression. The monthly climatic drivers T and P (X) were linearly detrended by month (Xd) and divided by their monthly standard deviation, resulting in monthly z-scores (Z).

$$Z_t = \frac{Xd - \overline{Xd}}{\sigma_{Xd}} \qquad (4)$$

For each pixel j, n (number of months: 18 for P and 12 for T) parameters were determined using linear regression. The primary challenge is to adequately characterize the "period of climatic influence." To address this, previous work considered time lags of up to 24 months for precipitation and 12 months for temperature (*7*). Given our shorter (20-year) record, we reduced the lag for precipitation to 18 months while retaining 12 months for temperature.

$$Y_j = b_{j1}Z_{j1} + b_{j2}Z_{j2} + \cdots + b_{jn}Z_{jn} + \varepsilon_j \qquad (5)$$

where Y is annual z-scores of GPP from 2001 through 2020, $b_{j(1\ to\ n)}$ represent regression parameters of monthly climatic influence on GPP annual z-score. The semi-annual time series (Xsa) contains the sum of the products of the original climate variables and the normalized absolute regression parameters:

$$Xs_{jt} = \sum_i \left(\frac{|b_{ji}|}{\sum_i b_{ji}}\right) X_{jit} \qquad (6)$$

where i represents the index of the months from 1 to 12 or 18, and t is years between 2001 and 2020. The monthly weights $\left(\frac{|b_{ji}|}{\sum_i b_{ji}}\right)$ represent the influence of the 12-18 months of climate variations on annual GPP variations.

By now, for each year, we integrated the impact of previous 12 months T and previous 18 months P for GPP anomalies extremes. For cumulative ET-P, as it is a cumulative variable, we use the last step of each yearly cumulative ET-P as that year's data.

## 6. Drought severity Classification

Basis: Σ(ET–P) deviations from climatological mean (Fig. S1)

Return period categories: The return period represents the statistical rarity of drought conditions, with shorter return periods (e.g., 5–10 years) indicating more frequent, near-normal dry conditions, and longer return periods (e.g., >100 years) corresponding to extreme drought events.

$$\text{Return period} = \frac{1}{P(Z > \frac{x - \mu}{\sigma})} \quad (7)$$

$$P\left(Z > \frac{x - \mu}{\sigma}\right) = 1 - \Phi(\frac{x - \mu}{\sigma}) \quad (8)$$

where $\Phi$ = standard normal CDF.

## 7. Spatial-Temporal Analysis

Spatial distributions: Mean annual median return periods (2001–2020 for Σ(ET–P); analogous for SSMI/GRACE-DSI).
Temporal trends: Daily/monthly global area fractions for each severity category; Linear trends (2001–2020) with significance ($p < 0.05$).
We calculated the annual median return period of drought indicators for each year at the global scale. These annual medians were then averaged over the whole period to produce a long-term mean return period maps shown in Fig. 1a for Σ(ET − P) (2001-2020), Fig. S2a for SSMI (2000-2020), and Fig. S3a for GRACE-DSI (2002-2020).
For analyzing the temporal change of global drought severity, we computed the daily/monthly global fraction of area for each return period category for the three indicators shown in Fig. 1b for Σ(ET – P), Fig. S2b for SSMI, and Fig. S3b for GRACE-DSI.

**Supplementary Text**

## *S1: Yearly cumulative ET-P and drought classification*

To detect drought events for each year, the cumulative difference between ET and P, Σ(ET-P), was calculated on a daily basis for each year separately from 2001 to 2020. Data from 2000 were excluded from this calculation because the record starts in March. Based on the annual Σ(ET-P), we computed the spatial means at both the global scale and for each continent individually (Fig. S2). Since we have data spanning 2001 to 2020, each region is represented by two subplots for clarity, each covering 10 years of data along with the overall mean and standard deviation calculated for the entire 20-year period. In Fig. S2, a given day is classified as a drought day when its Σ(ET-P) exceeds the multi-year Σ(ET-P) daily mean. If it exceeds the mean by more than 2 std, it is classified as a severe drought day.

## S2: Typical drought events

Numerous drought events occurred in recent years, including the severe drought in the Netherlands in 2018 and the drought in the Amazon during 2015-2016. We have chosen to analyze these two significant events primarily because of their severity and impact in different regions and climatic conditions. The drought event in the Netherlands in 2018 resulted in significant agricultural losses and water shortages, while the drought in the Amazon during 2015-2016 posed major threats to the ecosystem and biodiversity.

**Drought in NL 2018**

To compare with KNMI (Royal Netherlands Meteorological Institute) potential precipitation deficit $\Sigma(ET_0-P)$ between 1 April and 30 September of each year, we used data from 13 KNMI stations: De Bilt, De Kooy, Groningen, Heerde, Hoofddorp, Hoorn, Kerkwerve, Oudenbosch, Roermond, Ter Apel, West-Terschelling, Westdorpe en Winterswijk. We used precipitation data from these 13 KNMI rainfall gauge stations, and extracted reference evaporation ($ET_o$) data from thirty-six meteorological stations that are closest to the rainfall station and provide $ET_o$ data calculated from the Makkink method.

To calculate the cumulative difference of actual evapotranspiration and precipitation $\Sigma(ET-P)$, we used the same 13 KNMI rainfall gauge stations for the precipitation deficit for comparison, with precipitation from IMERG and ET from FluxHourly. Then we calculated spatial mean over the 13 stations for each year in 2000-2020 from 1 April to 30 September. We further calculated 21 years mean and standard deviation for comparison with the KNMI $\Sigma(ET_o-P)$. (Two stations lack some data: Station Oudenbosch does not have $ET_o$, station West-Terschelling does not have IMERG precipitation and ET).

From Fig. S8, the comparison between $\Sigma(ET-P)$ and $\Sigma(ET_o-P)$ highlights key differences in the characterization of drought condition, with $\Sigma(ET-P)$ providing a more realistic representation of land surface water availability. The KNMI-derived $\Sigma(ET_o-P)$ primarily reflects atmospheric water demand, which can exaggerate drought severity in years with high temperatures and radiation. For example, during the severe droughts of 2003 and 2018, $\Sigma(ET_o-P)$ exhibits a sharp increase, suggesting extreme water deficits. However, this metric does not account for the fact that actual evapotranspiration is constrained by soil moisture depletion. As a result, $\Sigma(ET_o-P)$ may overestimate drought severity by assuming that evapotranspiration would continue increasing as if water were available, while in reality, vegetation stress and soil moisture limitations suppress actual water loss. In contrast, Σ(ET-P), which incorporates actual water availability, provides a more reliable assessment of drought conditions. During 2003 and 2018, Σ(ET-P) remains lower, more accurately capturing the impact of severe soil moisture deficits on evapotranspiration. This better reflects the true water stress experienced by ecosystems and agricultural systems.

Focusing on the year 2018, a detailed examination of the spatial distribution of Σ(ET-P) from 1 April to 30 September reveals different drought dynamics in different regions in the Netherlands. Figure S9 illustrates the spatial distribution of Σ(ET-P) across the Netherlands over this period, showcasing the changes in water use throughout the growing season. At the end of April, the Σ(ET-P) values in the Netherlands were relatively low, indicating drought conditions were not severe. By the end of May, Σ(ET-P) began to increase in the northern and eastern regions, reaching around 60 mm, reflecting a growing demand for water by plants. By the end of June, with the exception of the southern province of Limburg, Σ(ET-P) values across other regions generally reached 80-100 mm, indicating a significant increase in plant evapotranspiration as temperatures rose. By the end of July, Σ(ET-P) values in most areas (except for Limburg and central regions) reached 140 mm, indicating strong water demand during the peak growing season. While these high values suggest considerable water use, Limburg and central areas exhibited better water conditions, indicating that drought severity was not as pronounced as in the northeastern region. However, by the end of August, Σ(ET-P) values along the western coastal regions began to decline, while eastern and northeastern regions maintained values around 140 mm, indicating that these areas faced greater water stress. By the end of September, the northeastern region continued to record Σ(ET-P) values at 140 mm, highlighting significant drought pressure experienced during the growing season. The spatial distribution shows that the northern and eastern regions maintained higher Σ(ET-P) values throughout the growing season, reflecting increased water demand and potential drought impacts. This spatial variability underscores the differing needs for water resource management and agricultural production in various regions.

The analysis of monthly SSMI values throughout 2018 revealed distinct temporal patterns in soil moisture dynamics (Fig. S10). From January to April, SSMI values were generally high, reflecting adequate soil moisture, with the exception of February, when reduced precipitation led to a noticeable decline. A pronounced drought period was observed from May to August, characterized by consistently low SSMI values, likely driven by elevated ET and limited rainfall. A slight recovery in soil moisture occurred in September, coinciding with a reduction in the cumulative (ET-P) values (Fig. S11). However, soil moisture declined again in October as (ET-P) values increased, indicating heightened water loss (Fig. S11). Recovery resumed in November, with SSMI values returning to near-normal levels by December, supported by increased precipitation and reduced ET.

**Drought in Amazon in 2015-2016**

The Amazon rainforest is vital to the global carbon cycle and has a major influence on Earth's climate. However, it's becoming more susceptible to prolonged droughts, worsened by climate change and human activities. One significant example of this vulnerability is the severe drought that occurred in the Amazon from 2015 to 2016 (*8, 9*). Cumulative (ET-P) from 1 Oct 2015 until 30 May 2016 was calculated for the Amazon basin (Fig. S12), when it experienced a significant drought event, which can be analyzed through the metric, $\Sigma(ET-P)$. During the initial phase from October 1 to 30, 2015, elevated $\Sigma(ET-P)$ values were observed in the eastern Amazon, indicating a deficit in moisture availability. The subsequent months from November to February demonstrated consistently high $\Sigma(ET-P)$ values, reflecting sustained water loss due to enhanced evapotranspiration relative to precipitation. This prolonged drought significantly impacted the ecosystem, leading to reduced plant growth, increased susceptibility to wildfires, and adverse effects on biodiversity (*10*). Notably, by March 2016, the spatial extent of high $\Sigma(ET-P)$ values began to contract, indicating a potential alleviation of drought conditions. By May 2016, $\Sigma(ET-P)$ values returned to baseline levels, likely associated with the onset of the wet season, which facilitated a recovery in soil moisture and ecosystem health.

To assess the drought's impact on carbon fixation, we computed monthly gross primary productivity (GPP) anomalies ($\Delta$GPP), defined as the difference between observed GPP and the 2001–2020 mean for each calendar month (Fig. S12b). Although $\Sigma(ET–P)$ had largely recovered by May 2016, $\Delta$GPP remained negative until January 2017, revealing a lagged photosynthetic response. Between October 2015 and January 2016, $\Delta$GPP anomalies were most severe— reaching more than –100 g C m$^{-2}$ month$^{-1}$ across northeast region. In February through May 2016, the magnitude of $\Delta$GPP declined somewhat, and the area of negative anomaly shifted westward. By June 2016 it expanded again and then gradually shrank, by September 2016 only edge regions continued to exhibit negative anomalies. $\Delta$GPP turned positive in most regions by January 2017, signifying a full recovery of carbon fixation. These coupled hydrological and physiological dynamics reveal that, although Σ(ET–P) returned to baseline by May 2016, ΔGPP remained negative for the rest of the year. Following several months of consistently high soil moisture, which promoted canopy regrowth, most photosynthetic capacity restored by January 2017.

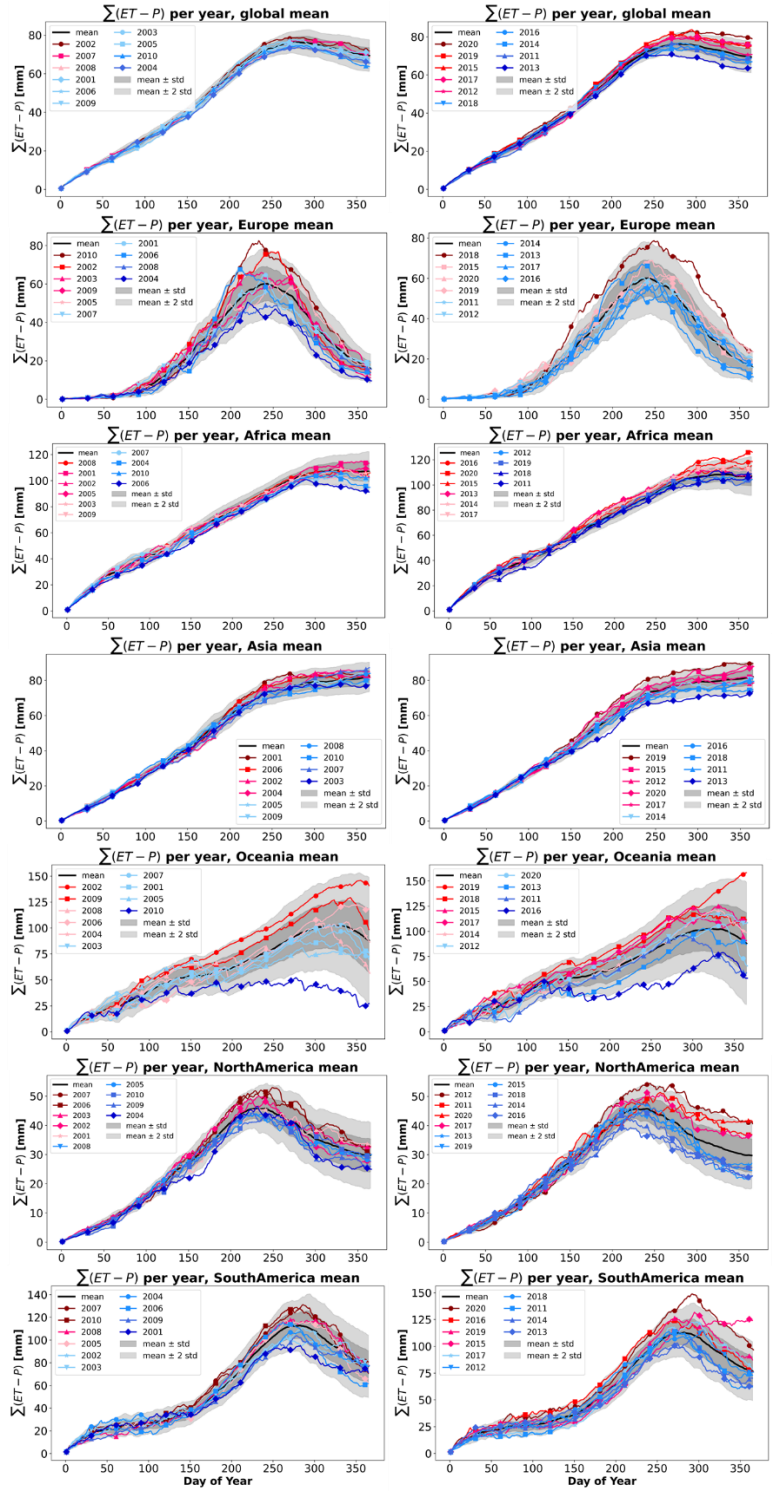

**Fig. S1.** Daily Σ(ET − P) per year, averaged over each continent, from 2001 to 2020. Each subplot shows two decades (2001–2010 and 2011–2020) with corresponding annual curves, multi-year mean, and ±2 standard deviation (shaded area). Red lines indicate dry years, while blue lines represent wet years.

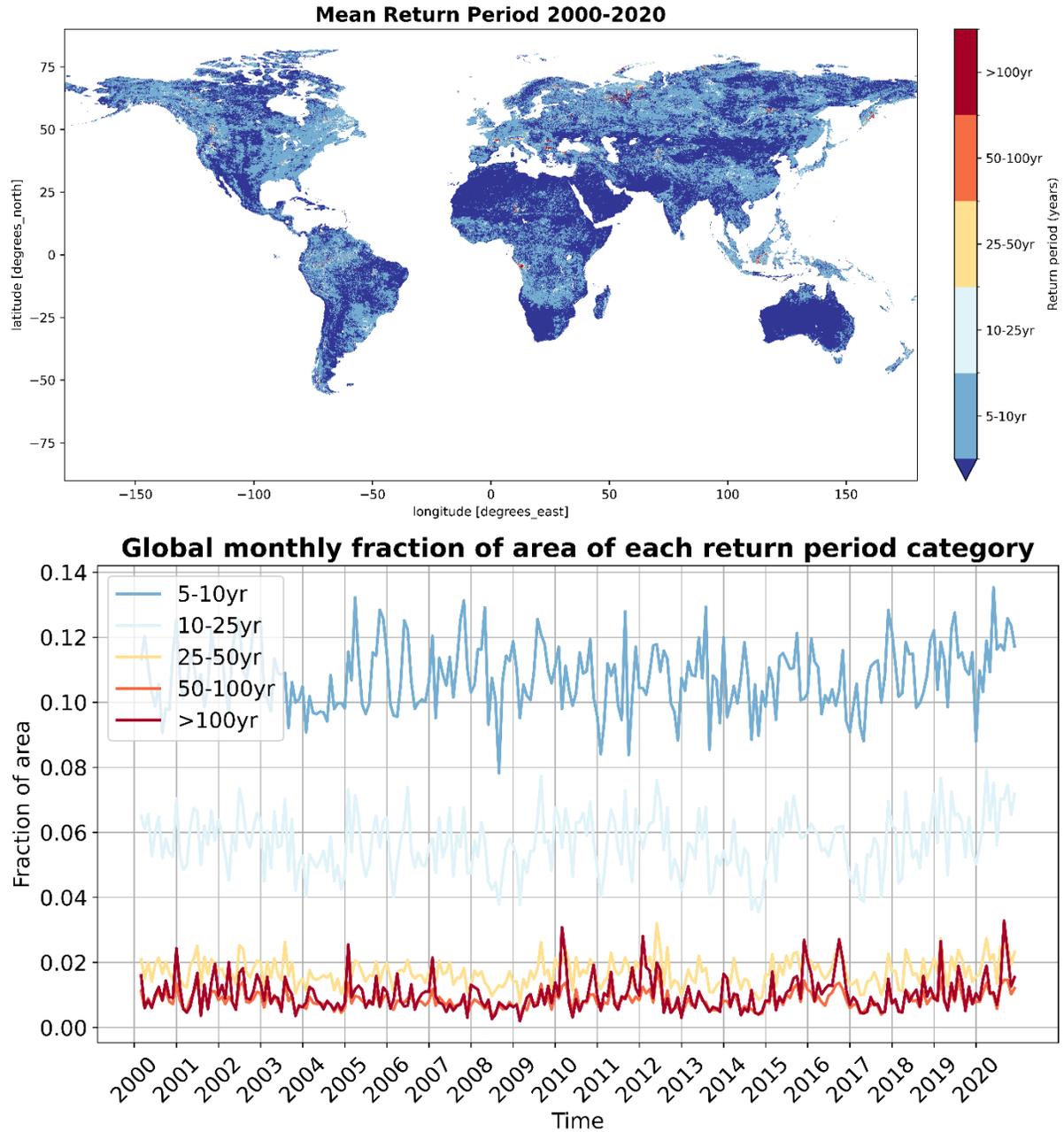

**Fig. S2.** (a) Global spatial distribution of mean drought return period based on monthly SSMI from 2000 to 2020 (b) Global monthly fraction of area for each return period category during 2000-2020.

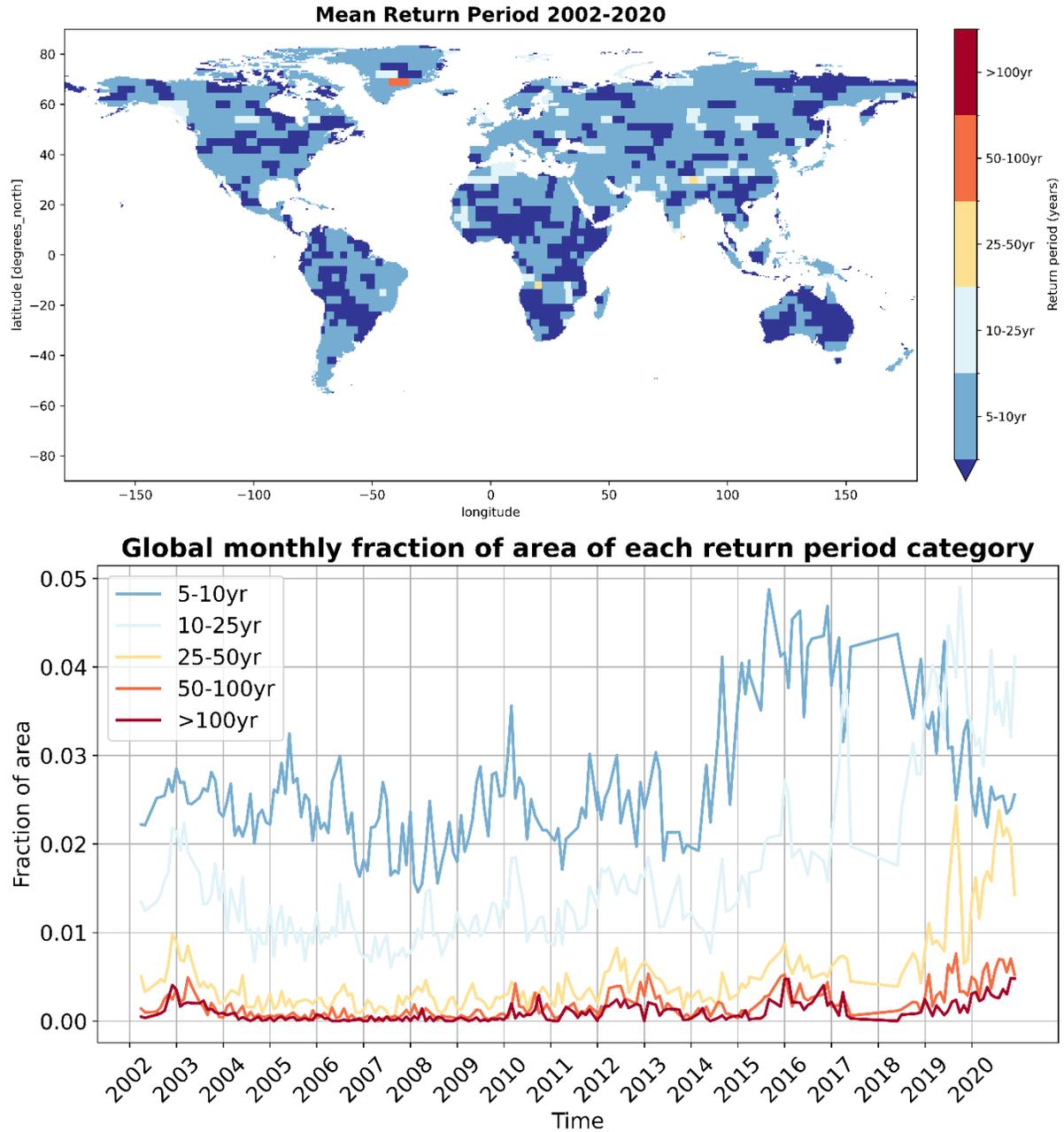

**Fig. S3.** (a) Global spatial distribution of mean drought return period based on monthly GRACE-DSI from 2002 to 2020 (b) Global monthly fraction of area for each return period category during 2002-2020.

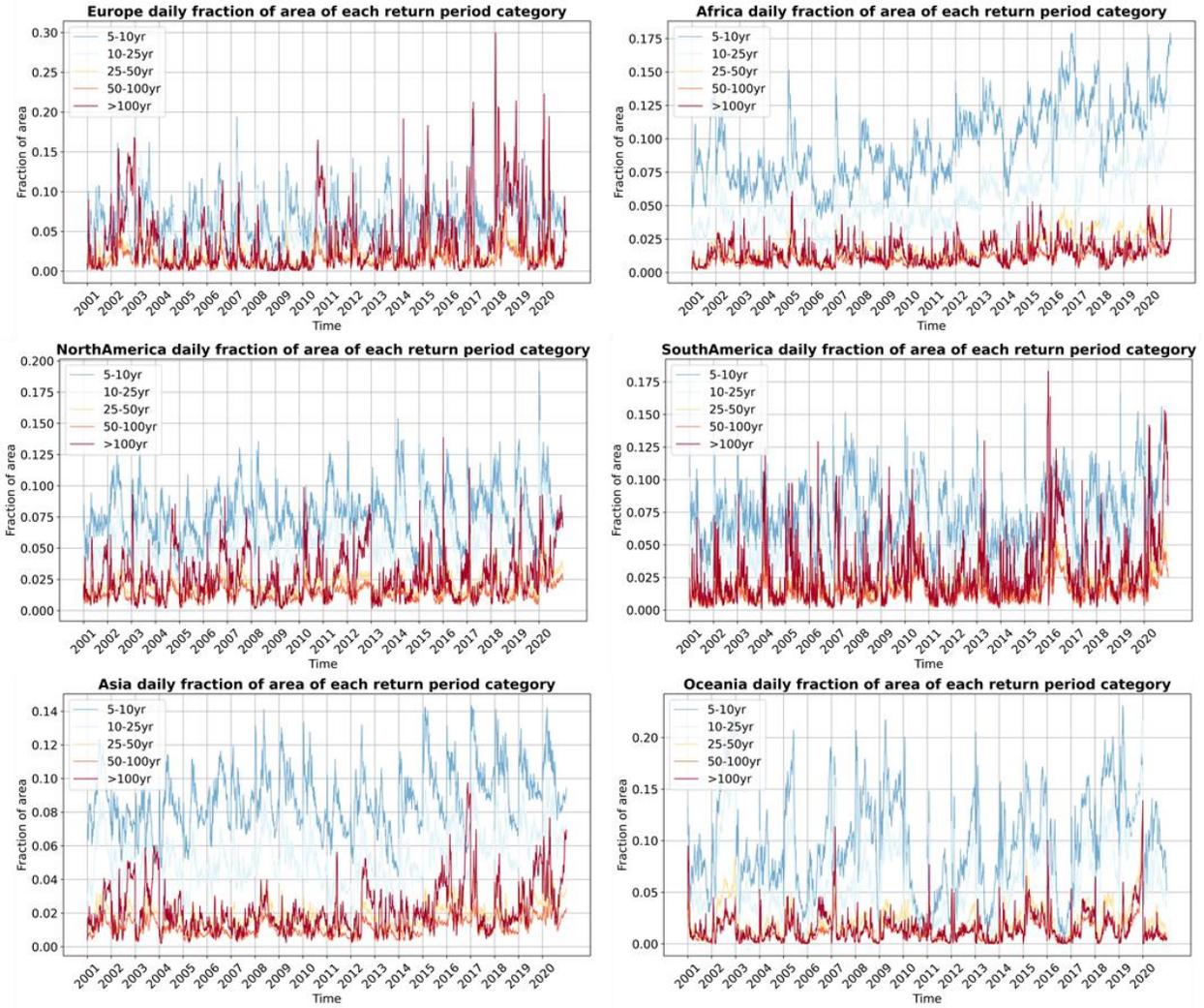

**Fig. S4.** Spatial extent of each severity category in 2001-2020 in each continent

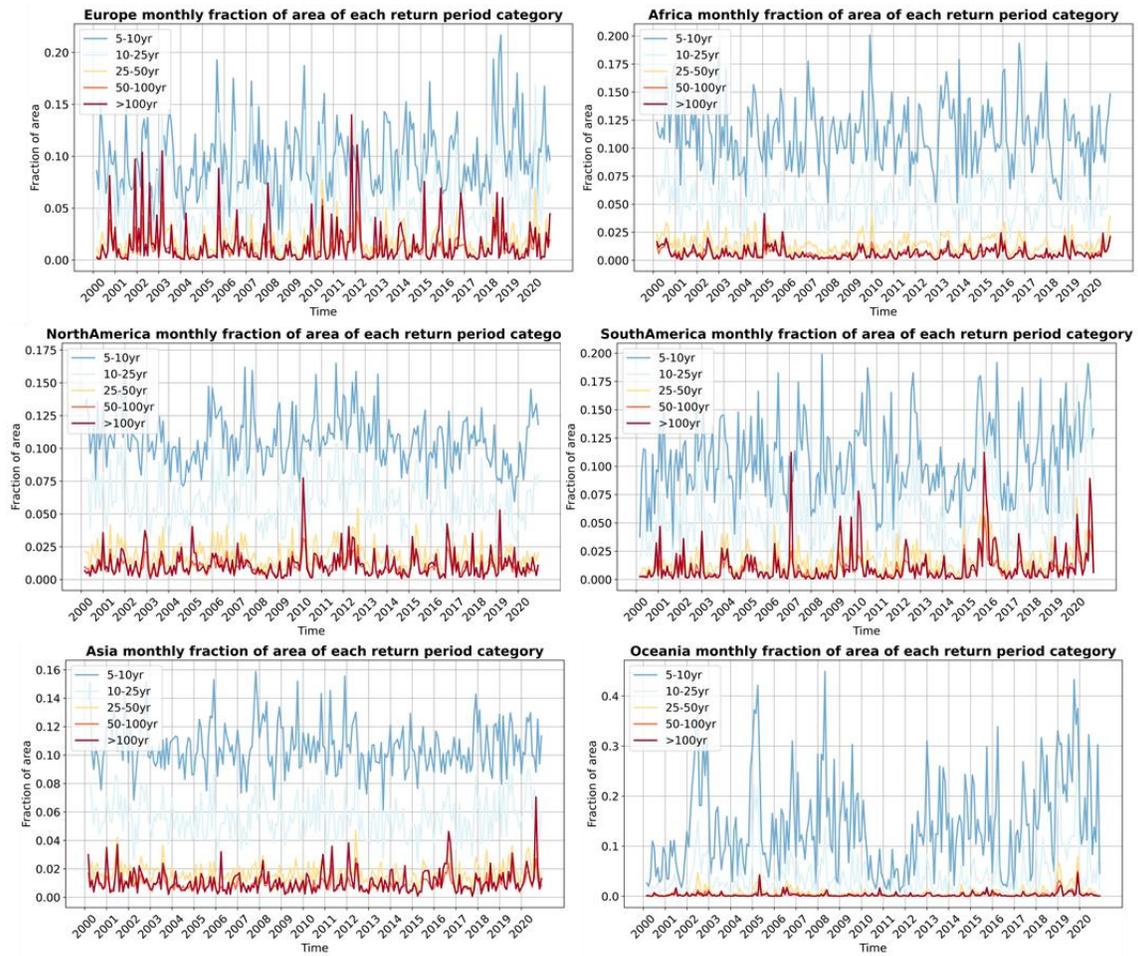

**Fig. S5.** Fraction of area of each return period category for drought events in 2000-2020 based on SSMI for each continent

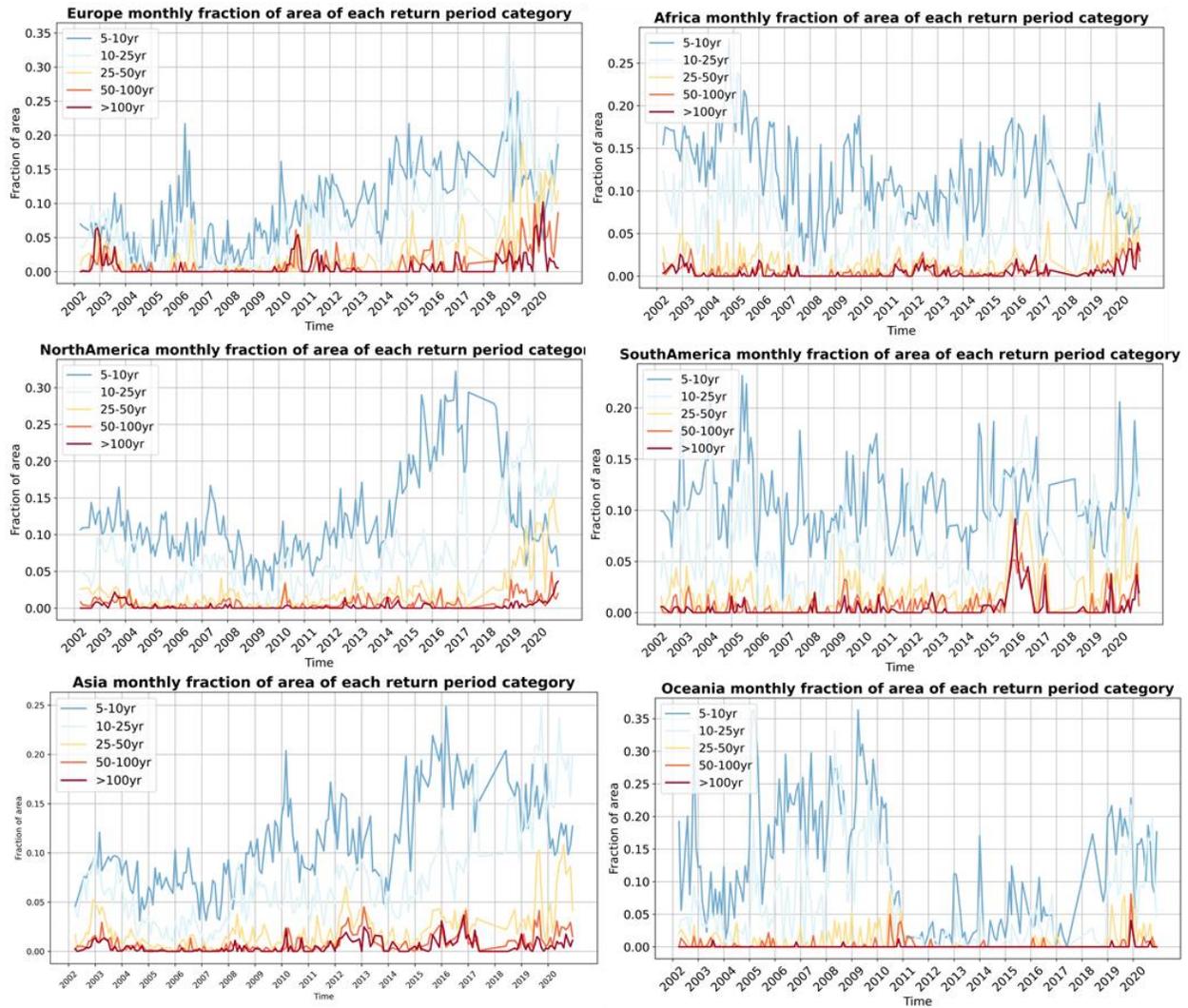

**Fig. S6.** Fraction of area of each return period category for drought events 2002-2020 based on GRACE for each continent

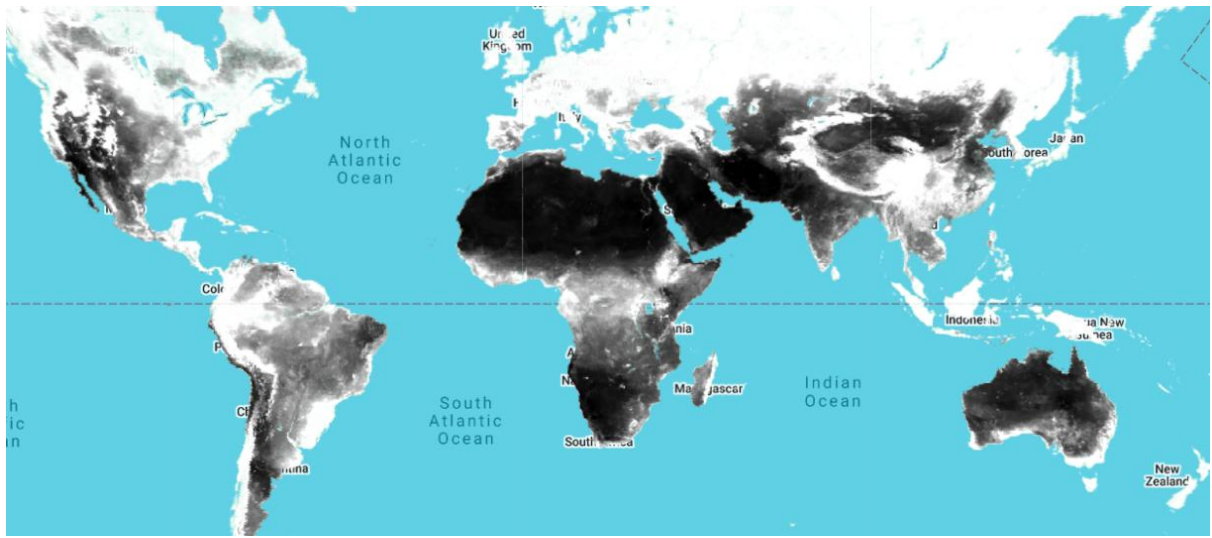

**Fig. S7.** Organic matter content from Soil Grids 250m (https://gee-community-catalog.org/projects/isric/)

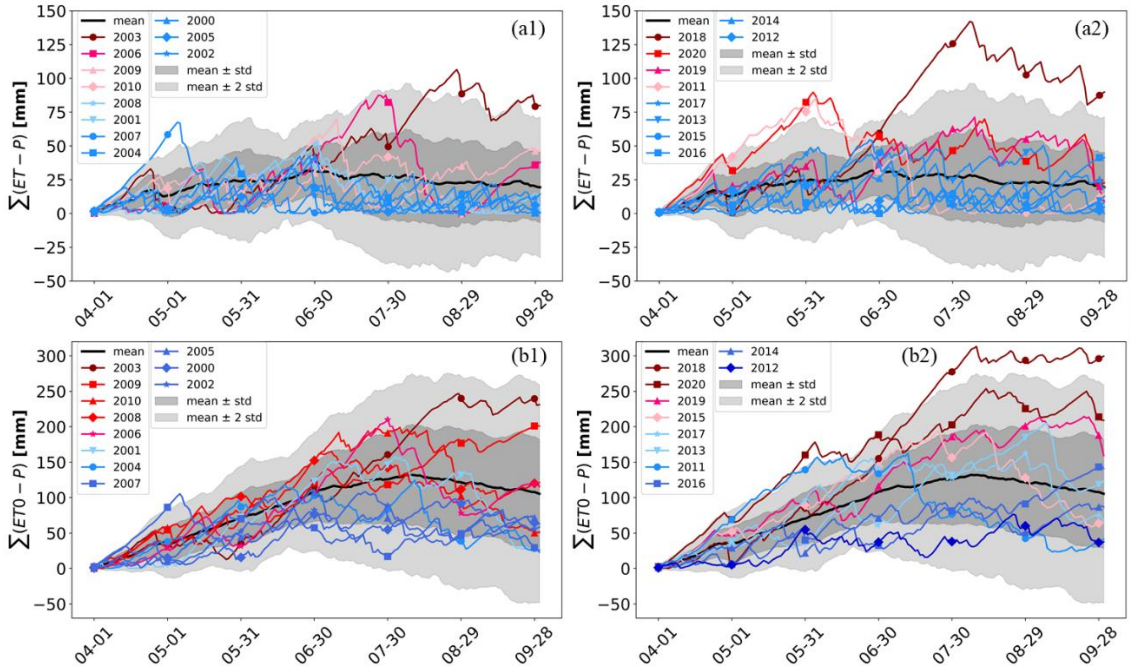

**Fig. S8.** a: cumulative (ET-P) in the Netherlands (a1: 2000-2010, a2: 2011-2020); b: cumulative (ET$_0$-P) (b1: 2000-2010, b2: 2011-2020).

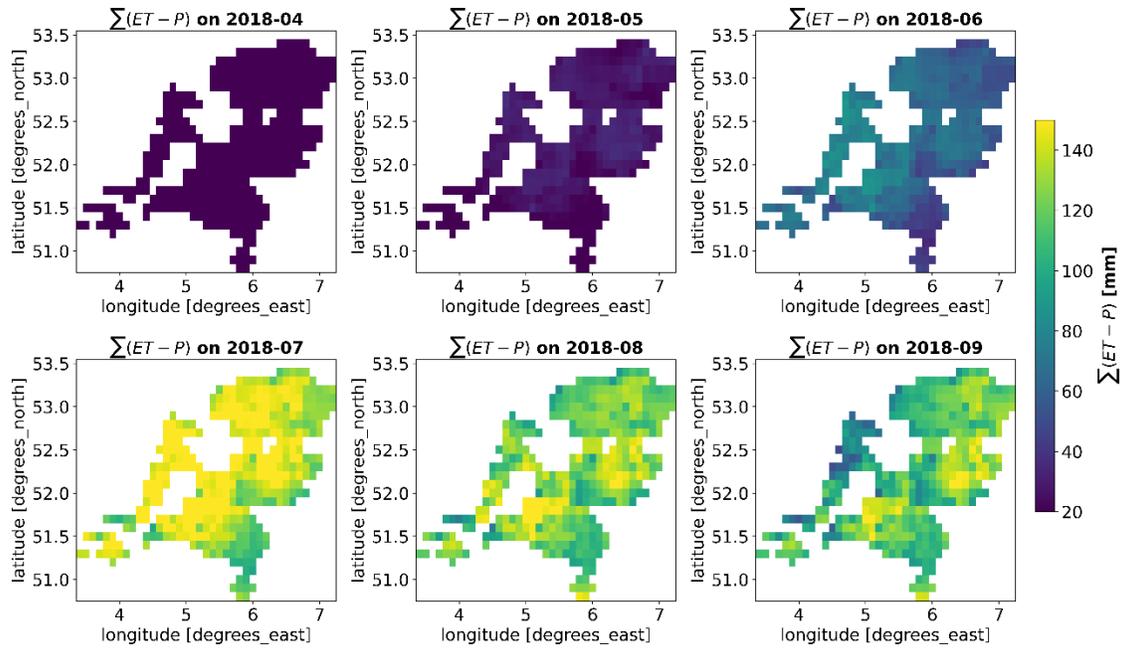

**Fig. S9.** Cumulative (ET-P) in the Netherlands in 2018 from April to September

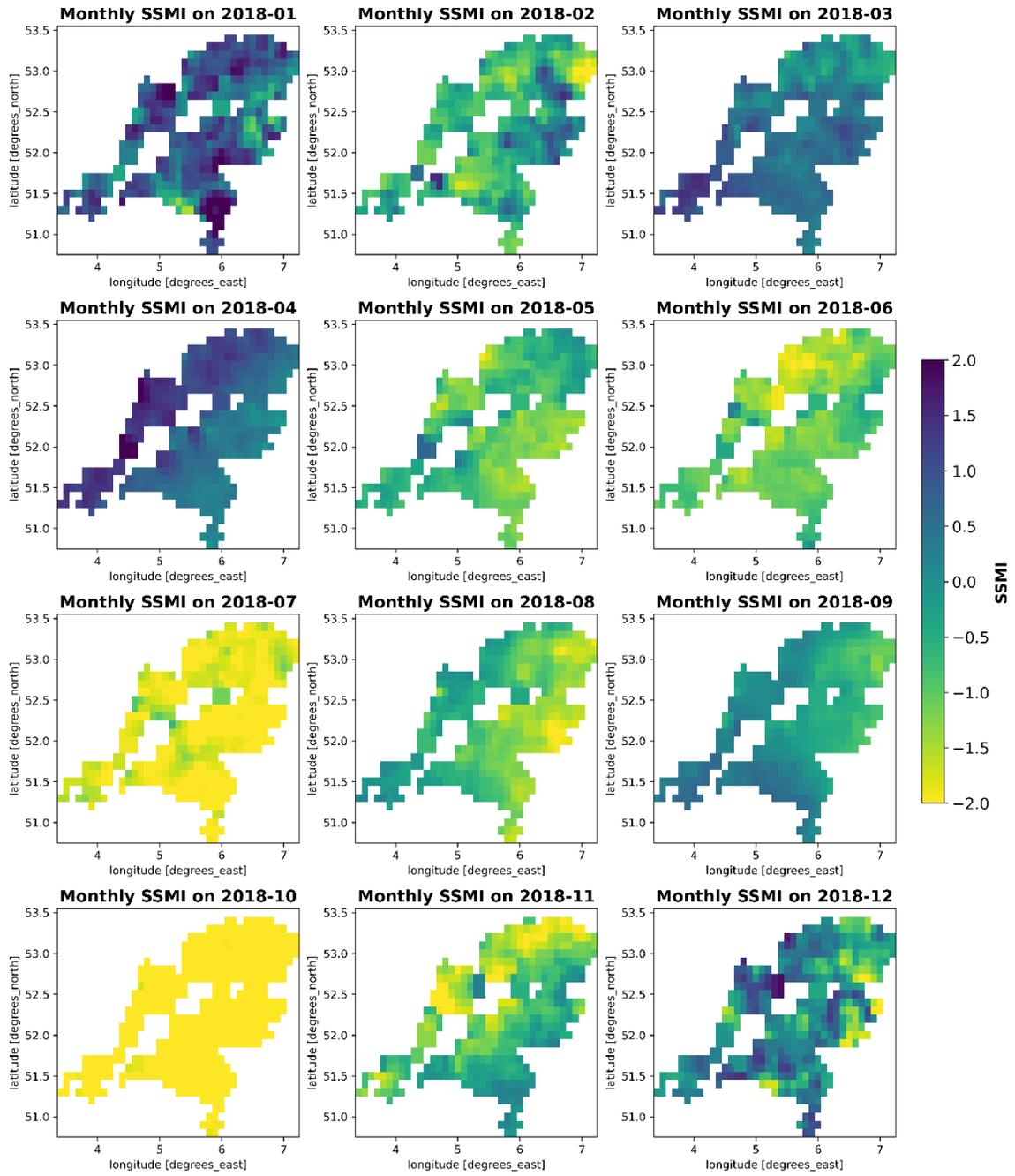

**Fig. S10.** SSMI in the Netherlands in 2018

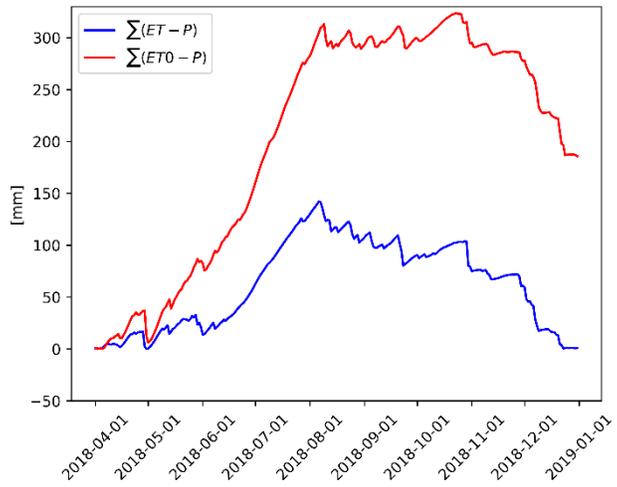

**Fig. S11.** Cumulative (ET$_0$-P) and cumulative (ET-P) in the Netherlands 13 stations in 2018 from April to December (same as Figs. S8 a4, b4, but extended to December)

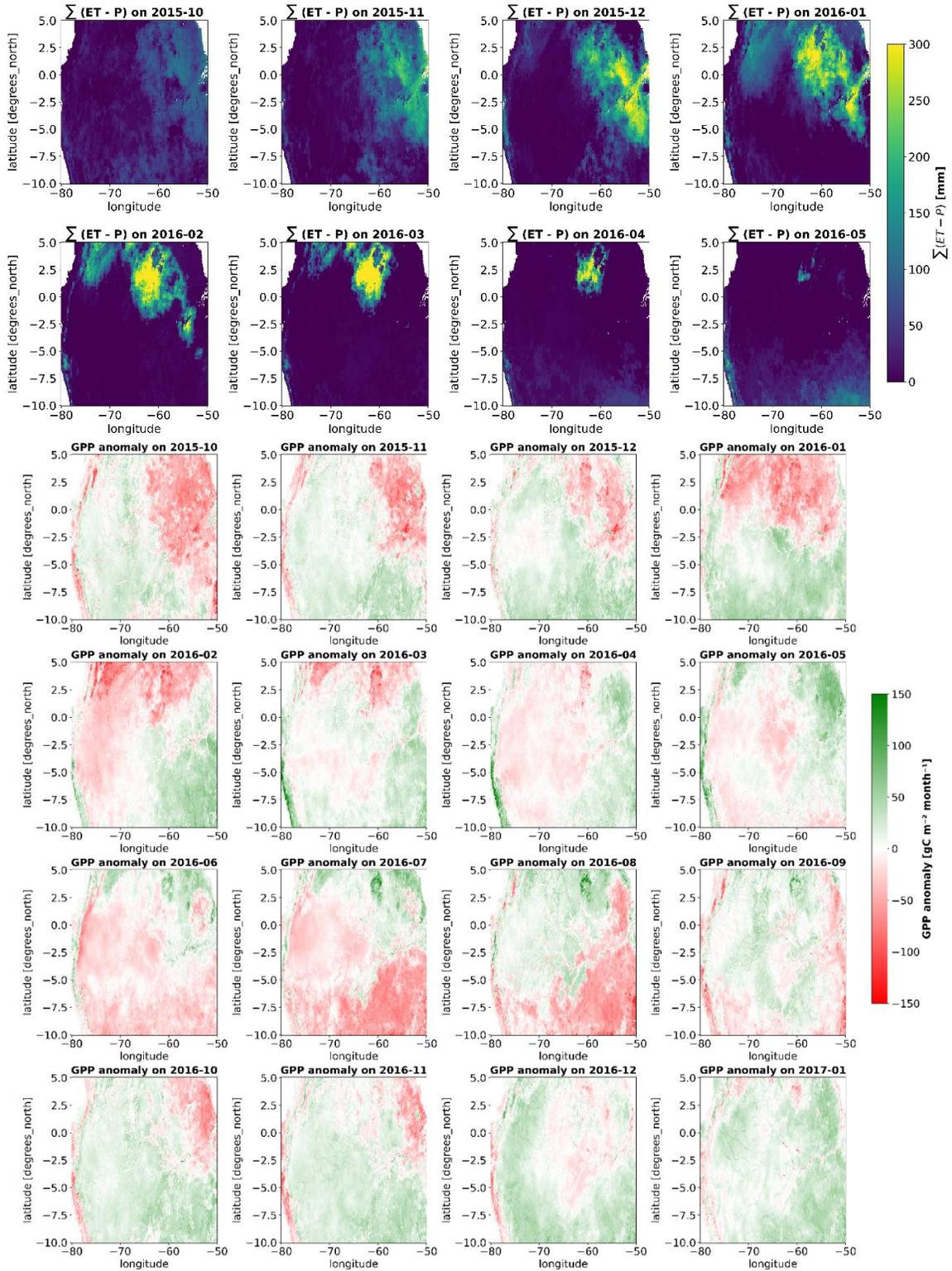

**Fig. S12.** (a) Cumulative (ET-P) in Amazon from Oct 2015 to May 2016. (b) GPP anomaly in Amazon from Oct 2015 to Jan 2017

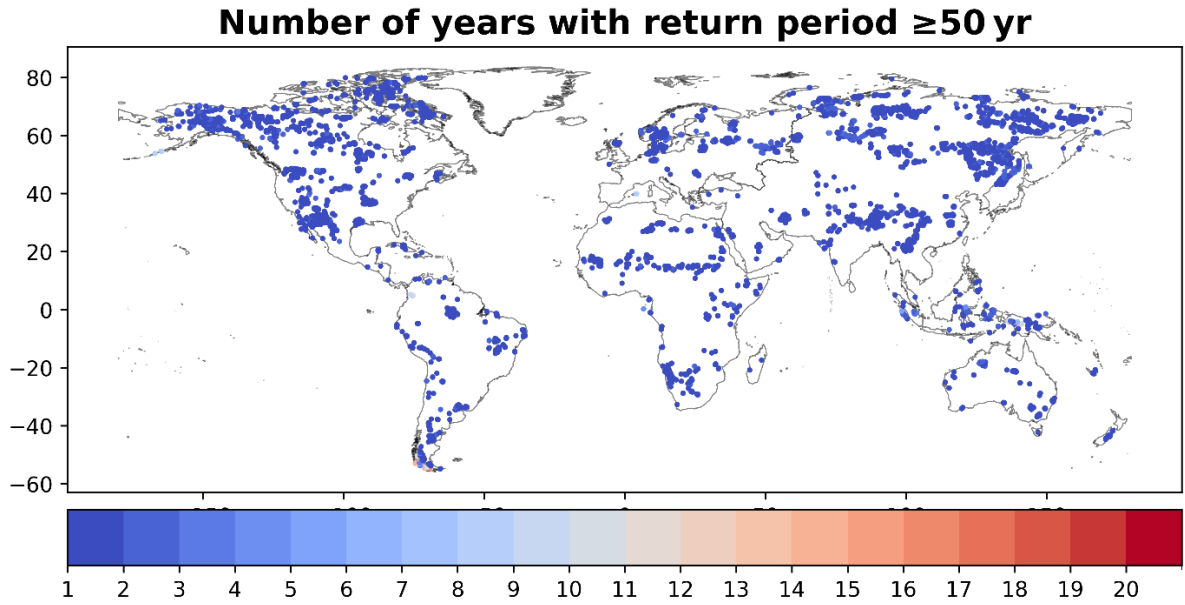

**Fig. S13.** In the hotspot regions (return period [100, +∞)), number of years with return period >= 50 yr.

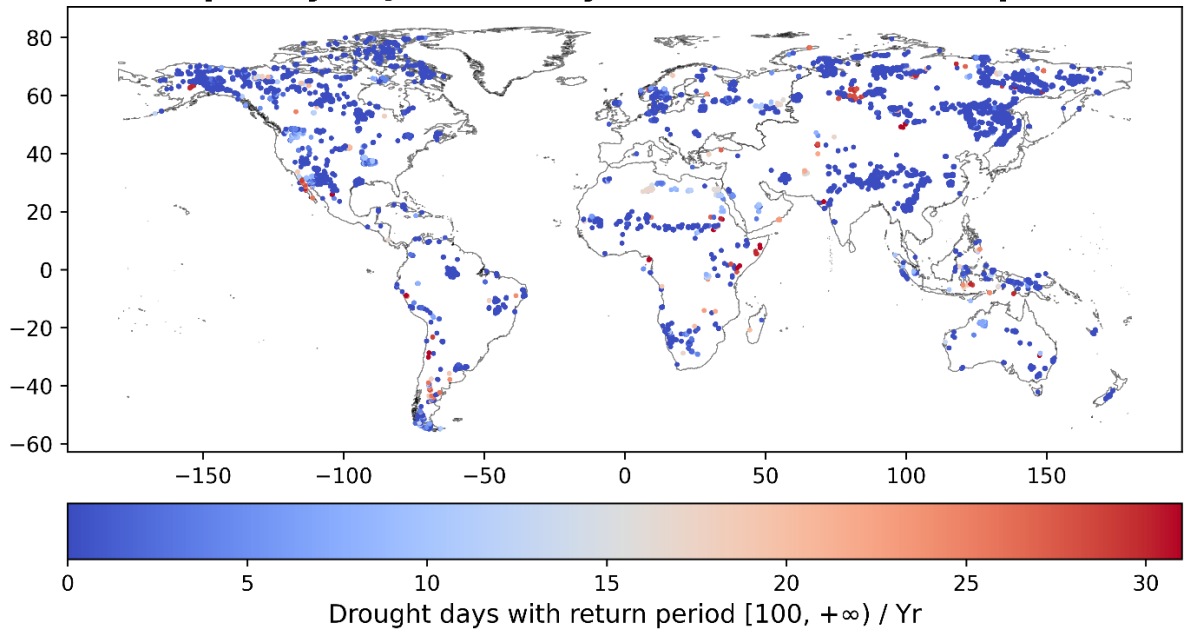
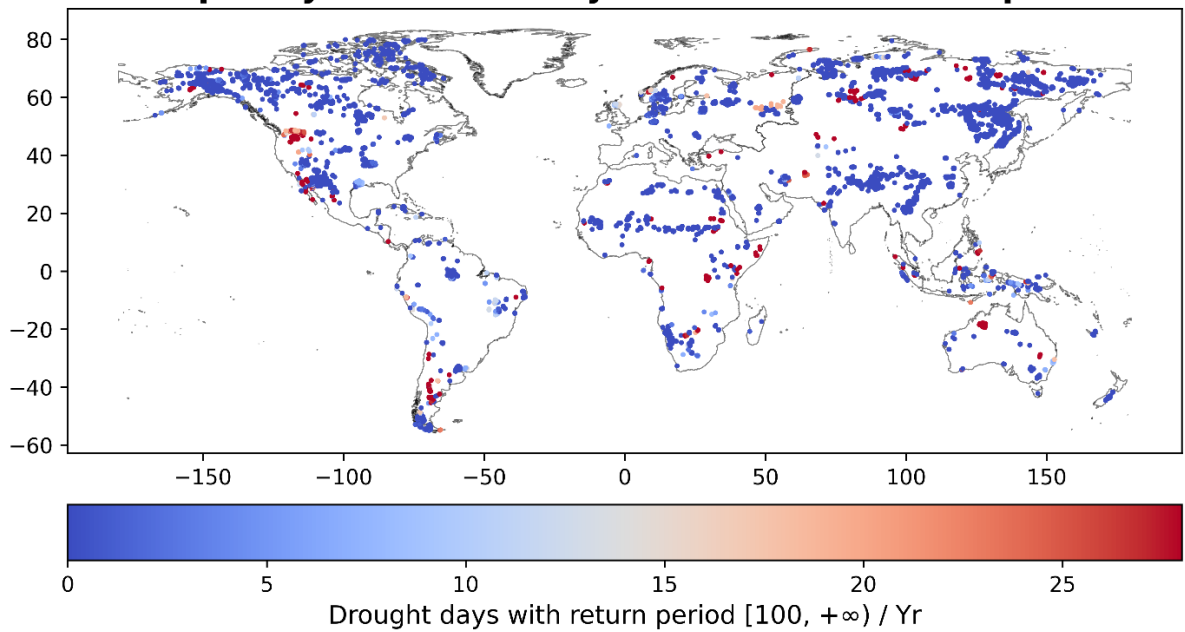

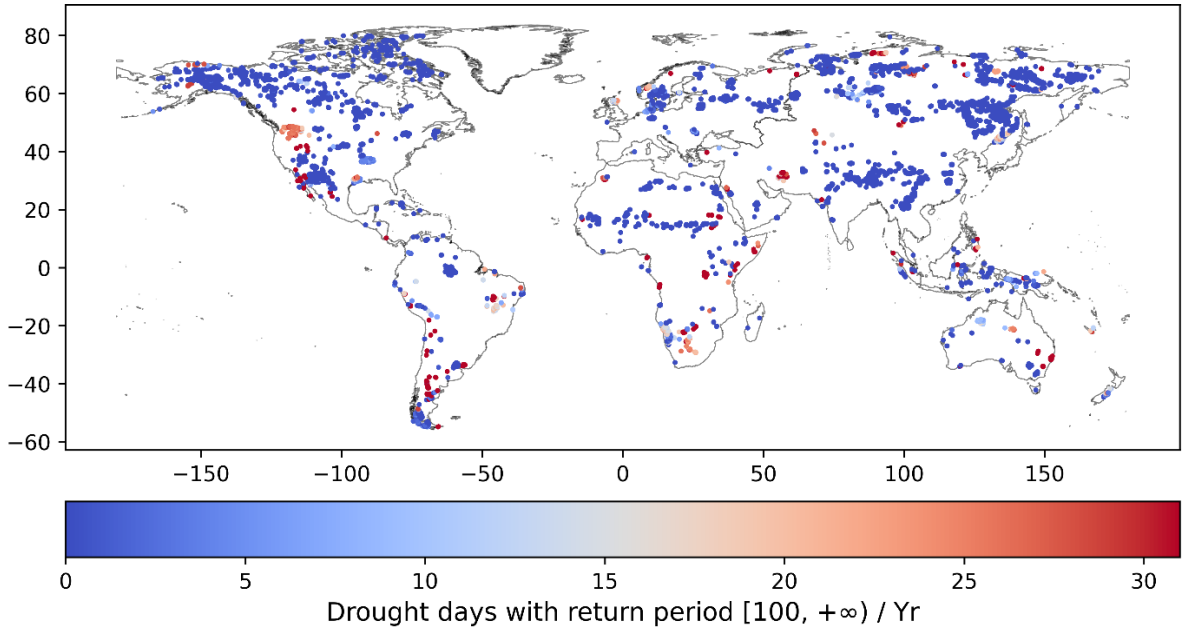

Frequency in Mar in the year with max return period

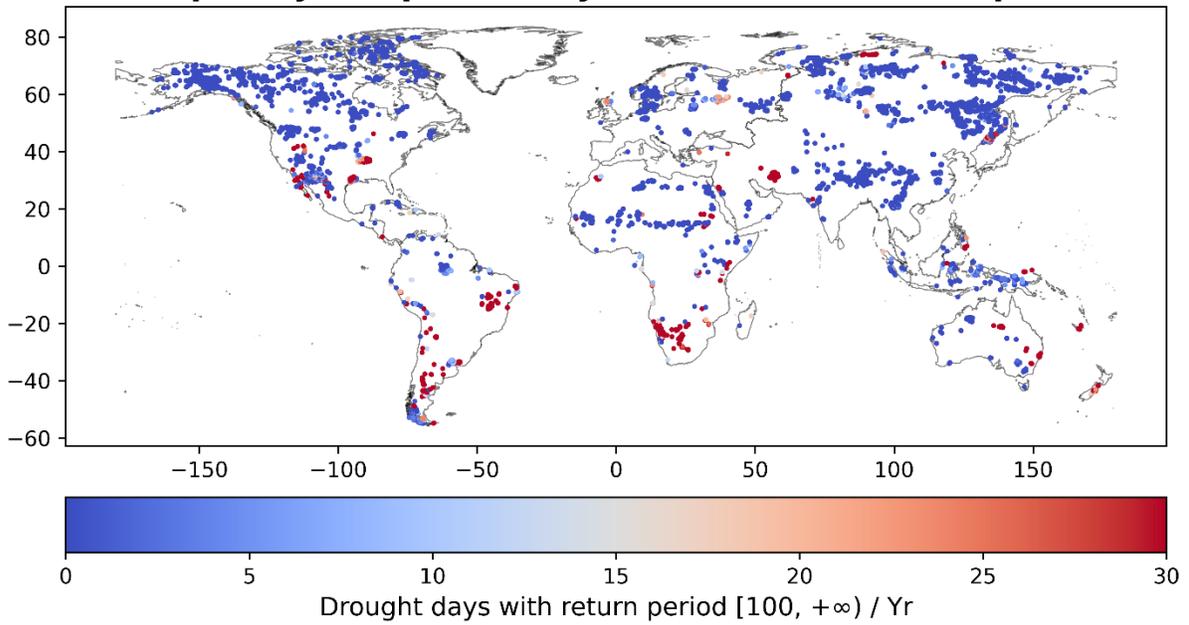

Frequency in Apr in the year with max return period

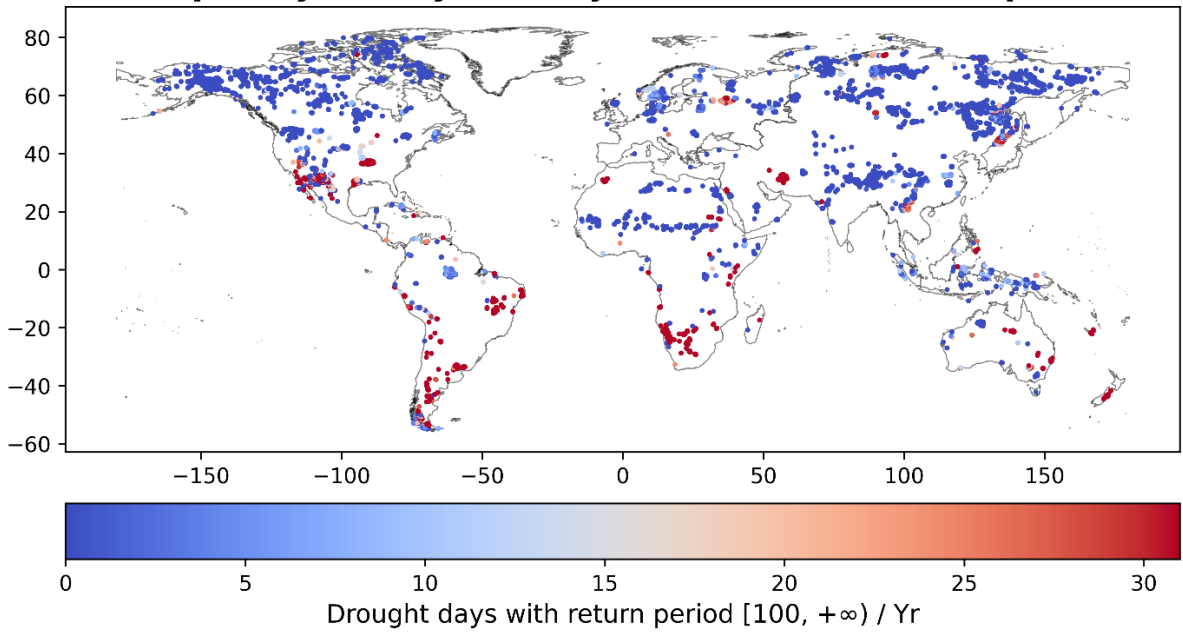
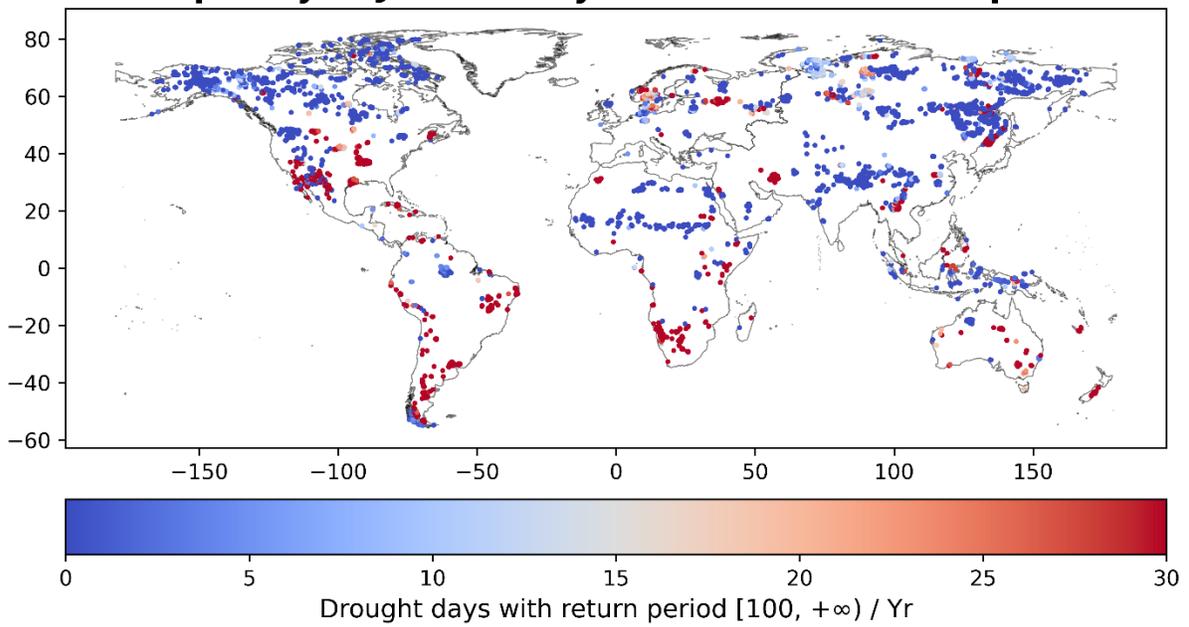

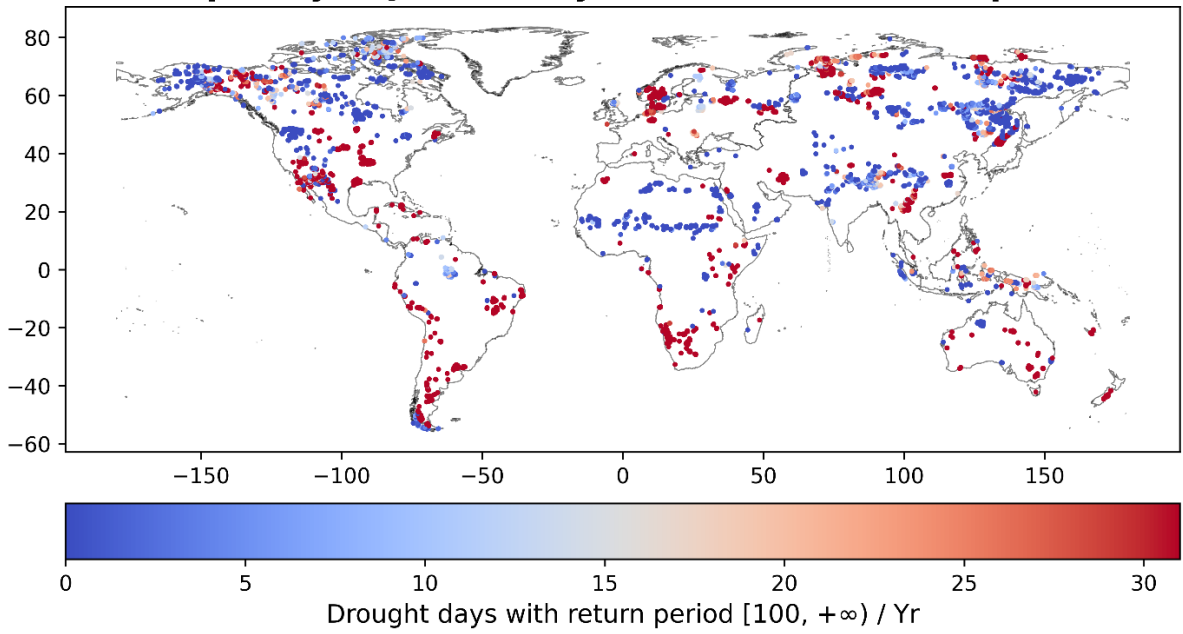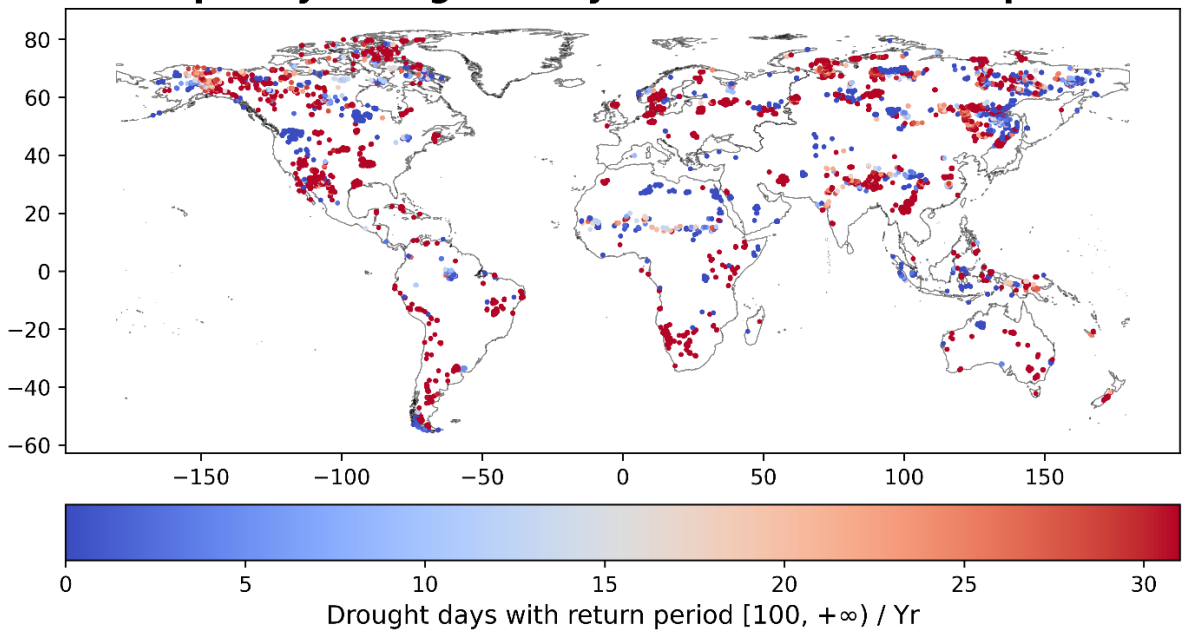17

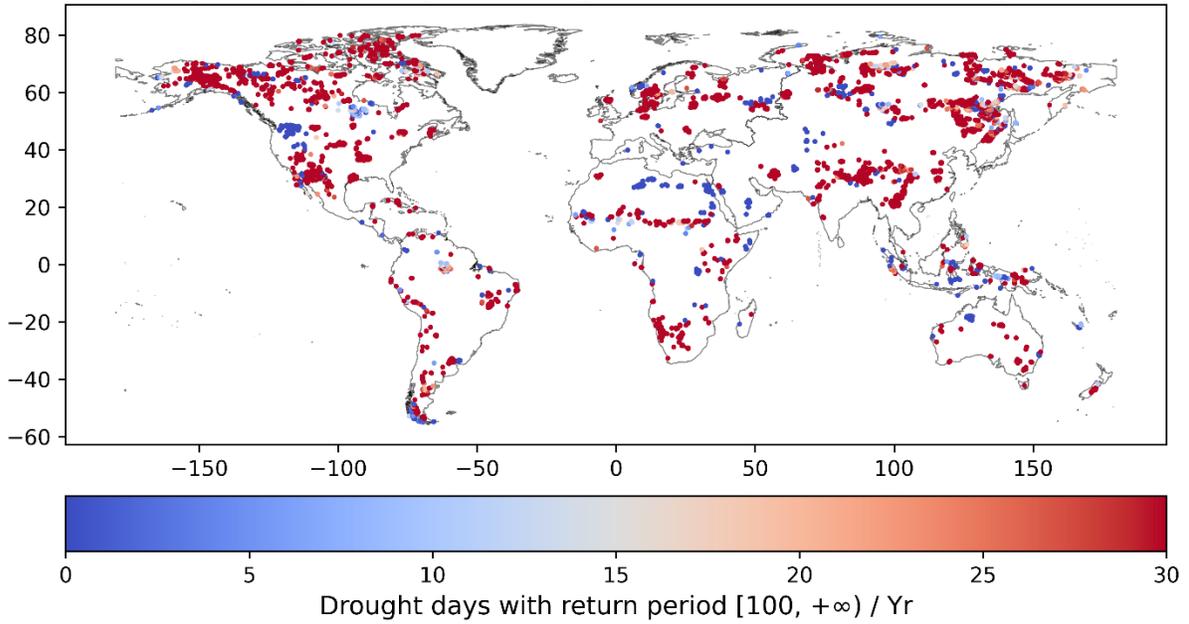
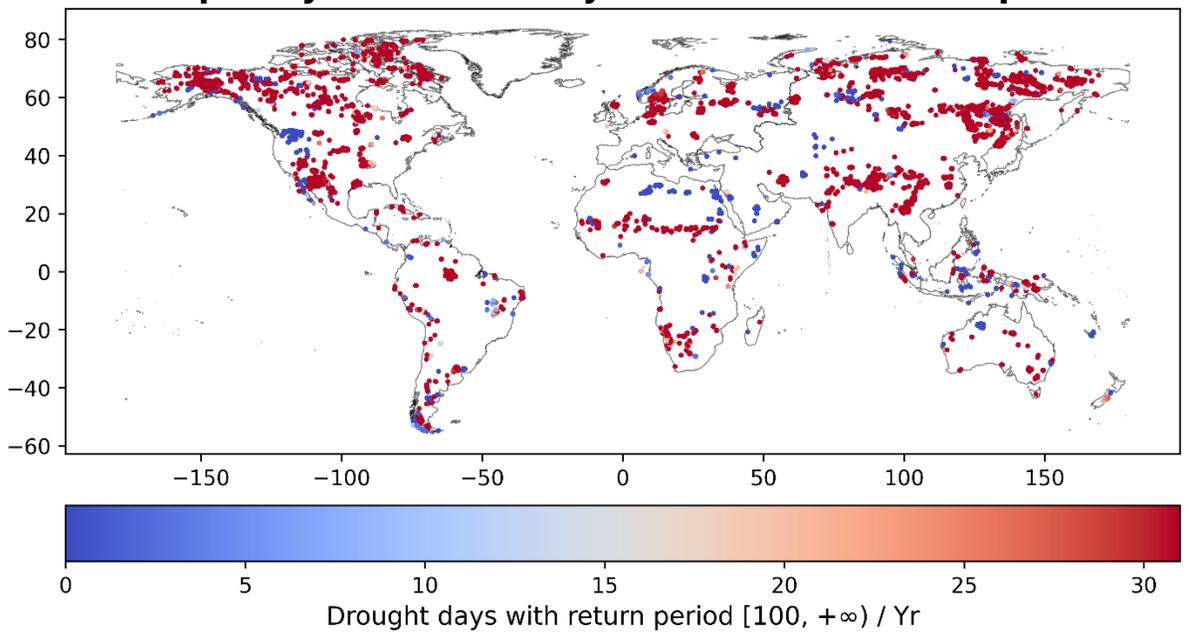

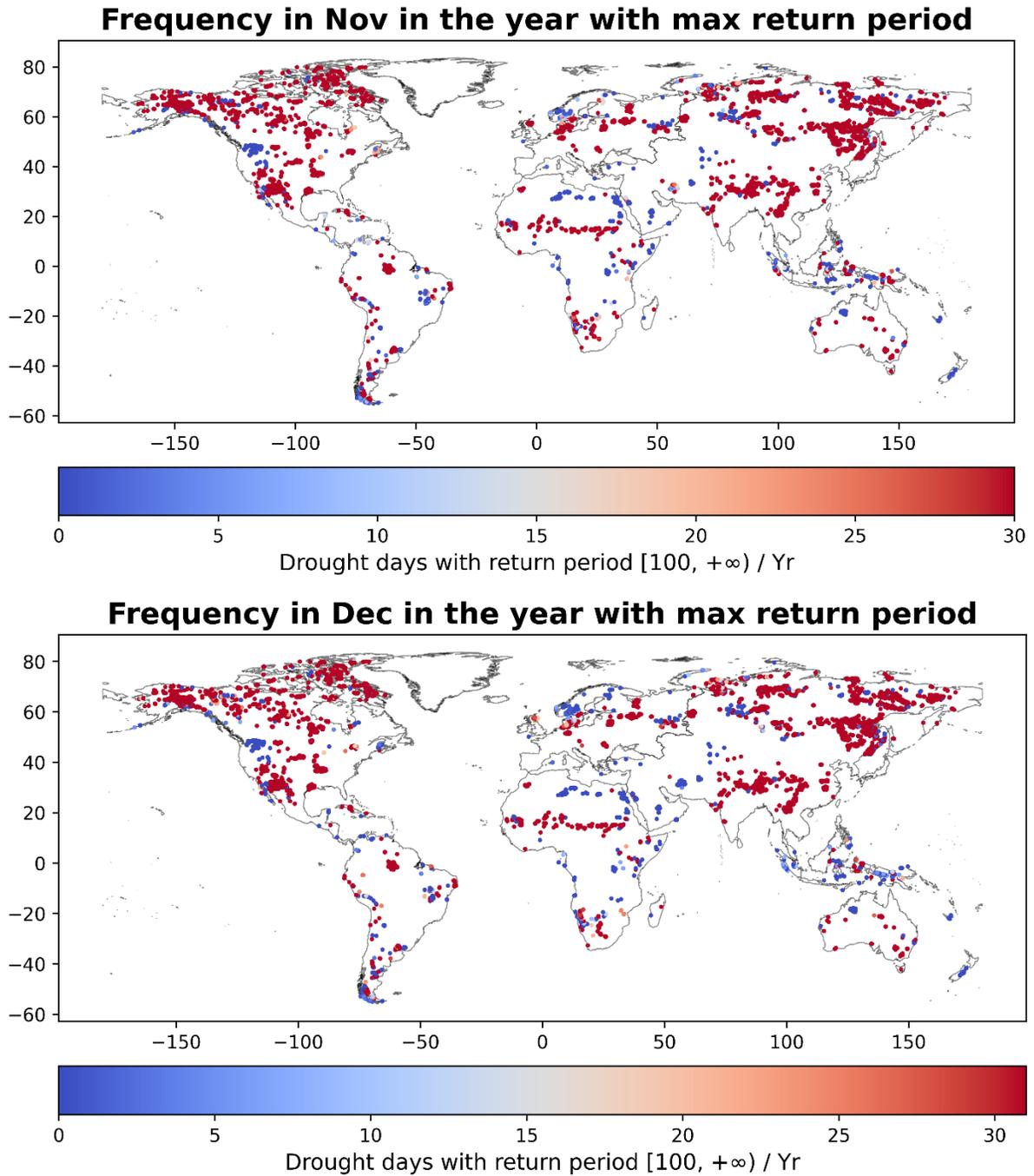

**Fig. S14.** The monthly frequency of category return period [100, +∞) in the year with max return period.

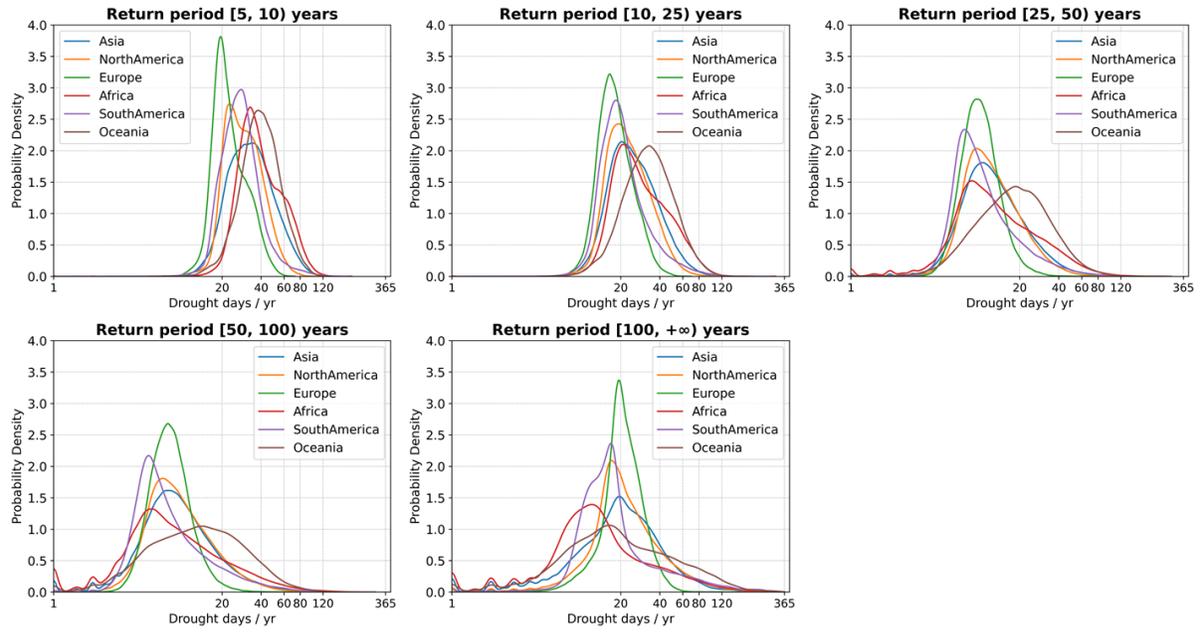

**Fig. S15.** PDFs of annual drought days frequency on log scale grouped by return period intervals: a) near normal [5,10) year, b) abnormally dry [10,25) year, c) moderately dry [25,50) year, d) severely dry [50,100) year, e) extremely dry [100, +∞) year.

**Table S1. Drought severity categories**

| Return period (year) | Value range | Category |
|---|---|---|
| [5,10) | $\mu+0.842\sigma \le F < \mu+1.282\sigma$ | Near normal |
| [10,25) | $\mu+1.282\sigma \le F < \mu+1.751\sigma$ | Abnormally dry |
| [25,50) | $\mu+1.751\sigma \le F < \mu+2.054\sigma$ | Moderately dry |
| [50,100) | $\mu+2.054\sigma \le F < \mu+2.326\sigma$ | Severely dry |
| [100, +∞) | $\mu+2.326\sigma \le F$ | Extremely dry |

**Table S2. Hotspot examples (T~Σ(ET–P) space)**

| Region | Climate Anomaly | Dominant Driver |
|---|---|---|
| Central Asia/NW China | Warm, dry | Water limitation (rainfall deficit) |
| Alaska/Scandinavia | Cold, wet | Energy limitation (low radiation) |
| Central/Eastern Europe | Cold, dry/wet | Season-dependent water/energy stress |

**Table S3. Divergence of metrics in extreme droughts**

| Indicator | Sensitivity | Extreme Drought Patterns | Hydrological Insight |
|---|---|---|---|
| Σ(ET–P) | Surface imbalance | Numerous hotspots; sharp temporal spikes | *Fast-onset meteorological droughts* |
| SSMI | Shallow soil moisture | Fewer extremes; smoother distribution | *Intermediate soil moisture response* |
| GRACE-DSI | Total water storage | Uniform; late-response peaks (2019–2020) | *Slow-propagating groundwater deficits* |